\documentclass[amsmath,amssymb,aps,prd,longbibliography,twocolumn,superscriptaddress,nofootinbib]{revtex4-1}

\usepackage[total={7.0in,9.5in}, top=1.0in, includefoot]{geometry}

\usepackage[usenames,dvipsnames,svgnames,table]{xcolor}
\usepackage{graphicx}
\usepackage{hyperref}
\usepackage{amsmath}
\hypersetup{
    hypertexnames=false,
    bookmarks=true,         % show bookmarks bar?
    unicode=false,          % non-Latin characters in Acrobat's bookmarks
    pdftoolbar=true,        % show Acrobat's toolbar?
    pdfmenubar=true,        % show Acrobat's menu?
    pdffitwindow=false,     % window fit to page when opened
    pdfstartview={FitH},    % fits the width of the page to the window
    pdfcreator={},   % creator of the document
    pdfproducer={}, % producer of the document
    pdfkeywords={} {} {}, % list of keywords
    pdfnewwindow=true,      % links in new window
    colorlinks=true,       % false: boxed links; true: colored links
    linkcolor=red,          % color of internal links
    citecolor=Green,        % color of links to bibliography
    filecolor=magenta,      % color of file links
    urlcolor=cyan           % color of external links
}

\begin{document}

\author{Samuel F. Way}
	\email{samuel.way@colorado.edu}
	\affiliation{Department of Computer Science, University of Colorado, Boulder, CO, USA}
	\affiliation{Spotify, New York, NY, USA}
\author{Santiago Gil}
	\email{santiago@spotify.com}
	\affiliation{Spotify, New York, NY, USA}
\author{Ian Anderson}
	\email{iananderson@spotify.com}
	\affiliation{Spotify, New York, NY, USA}
\author{Aaron Clauset}
	\email{aaron.clauset@colorado.edu}
	\affiliation{Department of Computer Science, University of Colorado, Boulder, CO, USA}
	\affiliation{Santa Fe Institute, Santa Fe, NM, USA}

\title{Environmental Changes and the Dynamics of Musical Identity}

\begin{abstract}
Musical tastes reflect our unique values and experiences, our relationships with others, and the places where we live. But as each of these things changes, do our tastes also change to reflect the present, or remain fixed, reflecting our past? Here, we investigate how where a person lives shapes their musical preferences, using geographic relocation to construct quasi-natural experiments that measure short- and long-term effects. Analyzing comprehensive data on over 16 million users on Spotify, we show that relocation within the United States has only a small impact on individuals' tastes, which remain more similar to those of their past environments. We then show that the age gap between a person and the music they consume indicates that adolescence, and likely their environment during these years, shapes their lifelong musical tastes. Our results demonstrate the robustness of individuals' musical identity, and shed new light on the development of preferences.
\end{abstract}

\maketitle

%*.*.*%.*.*.%
% INTRO 
%.*.*.%*.*.*%
Music is the soundtrack of our lives. It reflects our mood and personality, as well as the important people, places, and times in our past~\cite{denora2000music}. In this way, a person's \emph{musical identity}---the set of musical tastes or preferences that they hold, as well as anything that might modulate those preferences\footnote{As an example, the experience of being a classically trained musician affects the music that person is exposed to, how they evaluate it, and, ultimately, whether or not they like it.}~\cite{macdonald2002musical}---represents an ever-evolving depiction of their cumulative experiences and values. Understandably then, various scientific communities have devoted much attention to resolving what determines a person's musical tastes and, inversely, what can be inferred or predicted about someone based on their musical tastes. Progress in either direction broadens our understanding of the development of individual identity and culture, their rigidity and transmissibility, and the many roles that music plays in shaping our personal and social lives.

A common theme in musicology research explores individuals' use of music to modulate or express their mood, particularly among adolescents~\cite{north2000importance}, and the extent to which personality both shapes and is shaped by musical tastes~\cite{Schwartz2003,SCHAFER2017265}. Studies have found repeatedly that mood regulation is among the most common and important reasons for why people listen to music~\cite{sloboda2001emotions,saarikallio2007role}: music helps listeners relax, improve their mood, or simply relate to others through the emotions of music and its lyrics~\cite{Wells1991}. Music also plays a crucial role as a social currency, helping initiate and strengthen relationships~\cite{erickson1996culture}, for example, through the exchange of new music or shared experiences at live performances. In these ways, music brings together individuals, forming communities or ``scenes" around particular genres, artists, or the lifestyles they personify~\cite{bennett2004music,lena2012banding,cohen1991rock}.

When a community forms around some kind of music, the surrounding environment takes on an identity of its own. Cultural geographers have investigated this interaction between place and musical style~\cite{hudson2006regions,nash1996seven}, treating music as primary source material for understanding what places are or used to be like~\cite{kong1995popular}. Research in this direction has investigated, for example, the evolution of music styles in space and time~\cite{carney1974bluegrass}, the impact of tourism on shaping local musical culture~\cite{hebdige2003cutn,gibson2003bongo} and, the effects of migration on altering the musical landscape of places~\cite{carney1998music,Baily06}. In much the same way that a person's musical identity reflects important elements of their past and present experiences and values, the musical identity of a place tells the history of its people.

These studies highlight just a few of the broader categories of research on musical identity. Despite spanning a wide range of ideas and disciplines, a common theme emerges: musical identity---of individuals and places---is inherently dynamic and ever-changing. These changes happen both quickly, on the time-scale of our moods, and slowly, as the cultural landscape of our environments and music itself shifts gradually. However, many studies of individuals' musical identity analyze musical taste and its correlates at a single point in time. This limitation stems in large part from the difficulty of characterizing people's musical tastes, and tracking their changes over time. In recent years, though, more and more people listen to music online, providing a detailed digital record of how individuals' music consumption and tastes evolve over time, all over the world. 

In this study, we analyze music consumption patterns on Spotify, a popular music streaming platform\footnote{In mid-2018, Spotify reported having over 190 million active users worldwide, including more than 75 million users in the United States alone~\cite{spotify18}.}. We focus specifically on the United States, the world's largest market for music~\cite{ifpi2018}, and one of the earliest and largest adopters of online music streaming. Coincidentally, the U.S.\ is also one of the most studied locations in musicology research, providing rich context to guide our analyses and the interpretation of their results. We focus on understanding a key determinant in the development of individuals' musical identity: the role of environment in shaping a person's tastes. Specifically, we measure environments' effects on individuals' preferences by treating geographic relocation as the basis for constructing quasi-natural experiments, using a matched pairs experimental design to mitigate the effects of confounding variables and natural variation. In addition, we investigate the relationships between the age of a listener and the music they consume, informing the likely timing of when and where musical identity takes shape.

We begin by describing the primary sources of data used in our analyses, most notably individual music consumption histories and location summaries during three sample periods. We then outline our approach for characterizing musical taste profiles, built up from data-derived music genres, and for measuring whether changes in a listener's environment induce changes in their musical tastes. We conclude with a discussion of our results and an outlook on the future of musical identity research.

%*.*.*%.*.*.%*.*.*%.*.*.%
% DATA + METHODS
%.*.*.%*.*.*%.*.*.%*.*.*%

\section{Data and Methods} 
Our study analyzes the music consumption of Spotify users in the United States between December 2016 and February 2018. After excluding individuals with low activity, missing or invalid demographic information, or unreliable location data, our dataset spans the consumption histories of $N$=16,445,318 users, called ``listeners" throughout. Consumption histories include, for each listener: (1) daily stream totals for each artist, (2) daily stream totals for each song release year or vintage, and (3) state-level location data, estimated from the listener's streaming IP address.  In addition, these histories provide limited demographic information, including listeners' self-reported gender (coded as ``M", ``F", and ``X") and an estimate of their self-reported age, aggregated into 5-year windows (e.g., birth years between 2000--2004 and birth years between 1975--1979. These two examples represent the youngest and oldest age groups in our analyses). Aggregating daily histories, we constructed statistical profiles that summarize listeners' musical tastes and locations during several sample periods. We begin by describing our motivation behind selecting these time periods and, from them, formulating quasi-natural experiments. We then outline our method for characterizing individuals' musical tastes during these periods and analytical tools to measure the impact of geographic relocation on musical tastes.

\newpage

\textbf{Changes in listener environment.} In 2017--2018, over 32 million Americans relocated to a new residence, with approximately 4.8 million of those moves crossing state boundaries~\cite{us2018cps}. Past research has explored the regional subcultures of individual states, driven in large part by historical differences in the ethnoreligious identities, cultural preferences, and ways of life unique to the various groups who settled the United States~\cite{fischer1991albion,lieske1993regional}. In light of these regional differences, we consider the effects of environment on musical tastes, defining a listener's environment as their state of residence, which we infer from the person's most frequent streaming location.

Based on reports from U.S.\ moving companies~\cite{Allied2017}, the majority of state-to-state relocations happen during the summer months, when the weather is generally more convenient and most American education systems are on break. For this reason, we recorded state-level relocations between May and September 2017. Correspondingly, we defined a trio of three-month sample periods: one just before the moving months ($P_1$: March to May 2017); another immediately following the moving months ($P_2$: September to November 2017); and a third, several months later ($P_3$: December 2017 to February 2018). For each period, we aggregated users' artist streams and streaming locations over nine randomly selected days. We then identified individuals who relocated by noting changes in their most frequent streaming location between periods $P_1$ and $P_2$. In later sections, we will compare individuals' profiles during these periods to assess short-term effects of relocation.

To assess longer-term effects of relocation, we build on cultural norms in the United States specifying Thanksgiving and Christmas as travel holidays that are traditionally spent at home with family~\cite{benney1959christmas}. In 2017, an estimated 107 million Americans traveled in late-December alone, with about half of those trips exceeding fifty miles~\cite{aaa2017}. Our sample frame spans three such ``home holidays": Christmas 2016, Thanksgiving 2017, and Christmas 2017. Using streaming locations during these holidays (i.e., the five-day window centered around the holiday), we inferred plausible past locations for listeners. The results presented here consider listeners who spent two or more of these three holidays in a state other than their location during $P_1$ and $P_2$, suggesting a past move. Qualitatively, our results are unchanged for individuals who traveled for a single holiday. Naturally, this heuristic restricts our analyses to listeners who both observe and have the means to travel for these holidays~\cite{mallett2001long}. We discuss this limitation further in our conclusions but proceed with this important caveat in mind.

\medskip

\textbf{Characterizing musical tastes.}
People tend naturally to describe their musical tastes in terms of genres~\cite{rentfrow2003re}. This coarse-grained description highlights individuals' general tastes and masks details about regionally-specific artists within a particular genre. Given our goal of assessing changes in musical tastes, not predicting location, genres provide a suitably abstract characterization of tastes. But, genres can vary in size and specificity. Some studies suggest that there are as few as five dimensions to musical preferences~\cite{Rentfrow:2011aa}. In contrast, Spotify characterizes music using a growing list of over 1700 genres and subgenres~\cite{spotifygenres18}, ranging from broad categories like ``rock" and ``jazz" to narrow subgenres that distinguish, for example, ``metalcore" from ``power metal" and ``bebop" from ``hard bop." These differences in definitions present a challenge in choosing an appropriate level of categorization. 

\begin{figure}[!t]
	\centering
	\includegraphics[width=\linewidth]{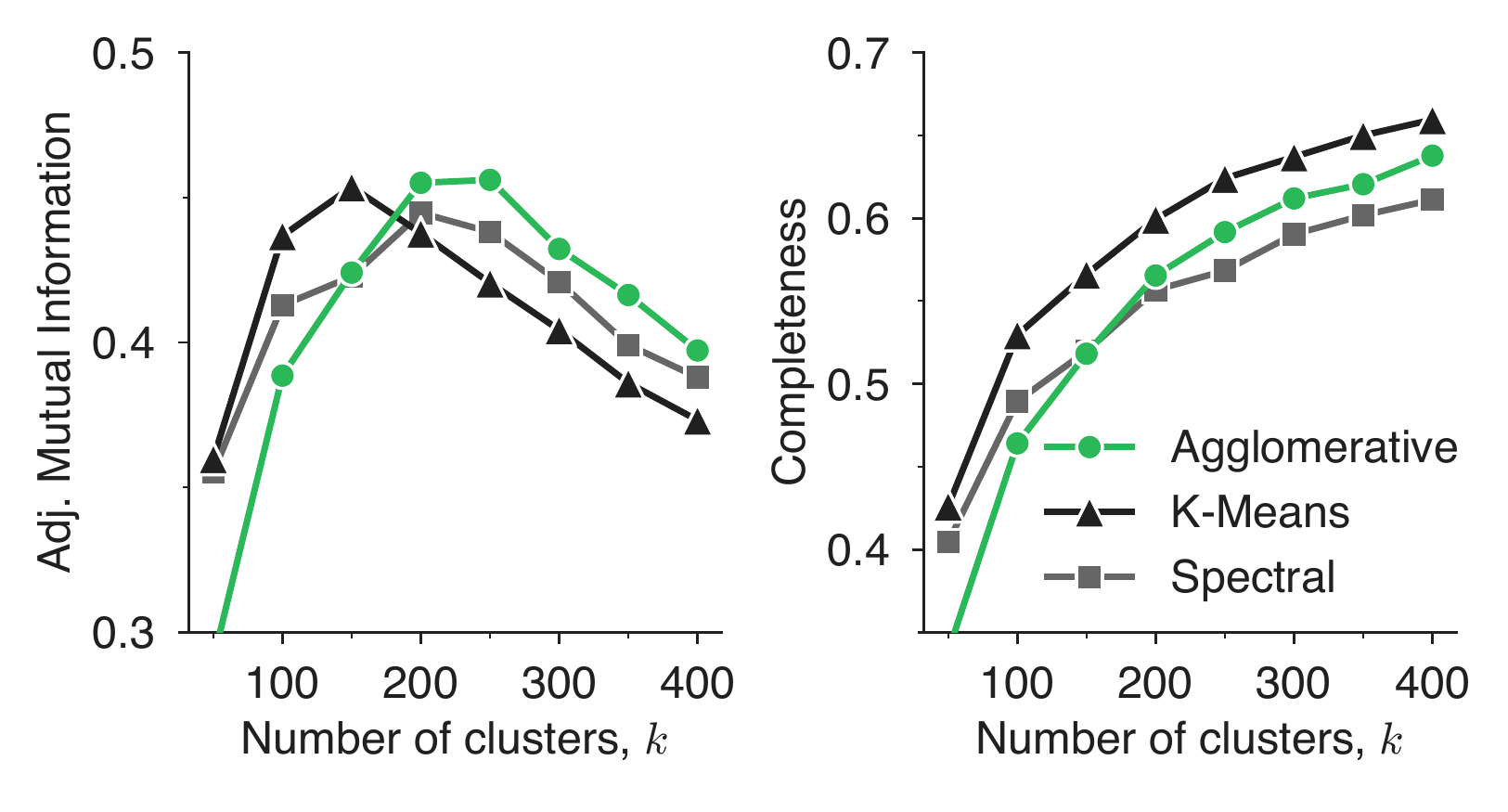}
	\caption{{\bf Clustering metrics suggest natural groupings of around 200 artists.} {\normalfont Average adjusted mutual information (left) and completeness (right) scores are shown, varying the number of clusters in three clustering methods. Adjusted mutual information peaks, and completeness begins to level out at around 200 artists, suggesting a reasonable number of genres for our analyses. Other approaches, including silhouette analysis and information criterion methods (not shown) further support this number.}}
	\label{fig:clusterscore}
\end{figure}

Here, we adopt a data-driven approach, defining genres as clusters of artists, whose similarity is derived from the frequency with which listeners stream two artists in succession. We focus on the $N\!=\!10,000$ most-streamed artists in the U.S., who collectively account for the vast majority of all streams in the country. For these artists, we construct a transition matrix $T$ whose entries $T_{i,j}$ denote the probability that a listener streamed a song by artist $i$ then artist $j$ during our sample frame. We then converted $T$ into an $N\!\times\!N$ distance matrix, $D$, by computing the pairwise correlation distances between each pair of artist vectors $T_i$ and $T_j$:

\begin{equation}
\label{eqn:corrdiff}
D_{i,j} = 1 - {(T_i-\bar{T_i}) \cdot (T_j-\bar{T_j})\over{||T_i-\bar{T_i}||_2||T_j-\bar{T_j}||_2}},
\end{equation}

\noindent where $\bar{T_i}$ is the mean of the elements of vector $T_i$, and $||\cdot||_2$ is the Euclidean norm.

\begin{figure}[!t]
	\centering
	\includegraphics[width=\linewidth]{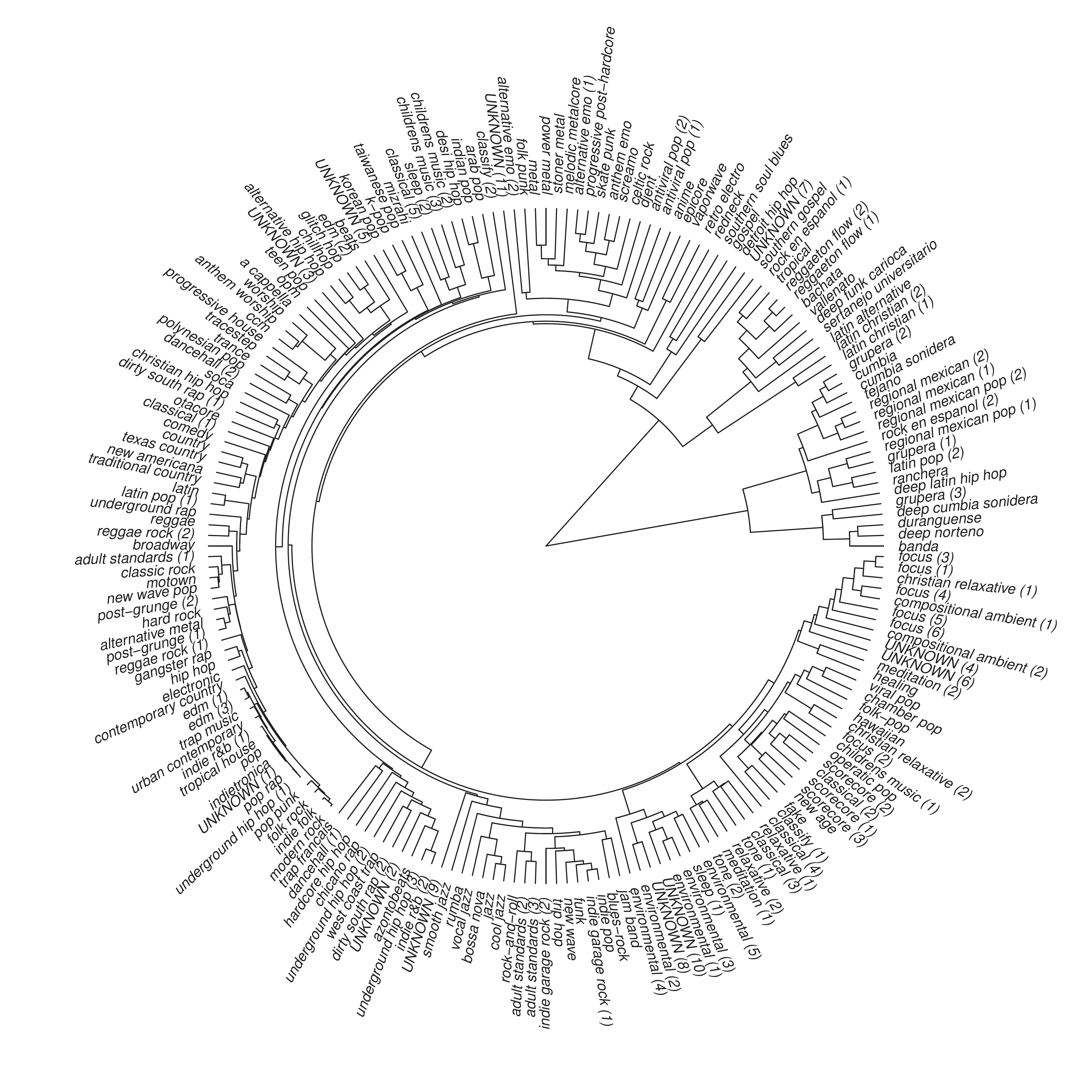}
	\caption{{\bf Dendrogram showing relatedness of the 200 data-derived genres.} {\normalfont Constructed using the UPGMA algorithm and genre-genre correlation distances, this dendrogram serves as the basis for our UniFrac-based comparisons of musical taste profiles. This clustering captures many known relationships between genres. Notably, the largest distinction in the tree, shown near 2 o'clock, splits musical genres by the language of their lyrics. Genres titled ``UNKNOWN" represent groups of artists with no specified genre labels on Spotify. Numbered genres differentiate individual clusters having the same most-common Spotify genre label.}}
	\label{fig:dendrogram}
\end{figure}

We then evaluated several unsupervised clustering algorithms---agglomerative, k-means, and spectral clustering---to obtain data-driven clusters of similar artists, or genres. To determine an appropriate number of clusters, we calculated cluster purity metrics over a varying number of clusters. Qualitatively, the three clustering techniques produce similar-scoring partitions of the data, and suggest a natural number of between 150 and 250 genres of music (Figure~\ref{fig:clusterscore}). Based on this analysis, we characterized listeners' musical tastes using $K\!=\!200$ genres, derived from the agglomerative clustering results. Past studies suggest that this level of abstraction may provide more detail than is often described or even perceived by typical listeners~\cite{rentfrow2003re}. Our characterization of musical tastes thus implicitly assumes that listeners are attuned to differences between these 200 genres, making our analyses perhaps more sensitive to change than listeners themselves. Finally, to name these genres, we selected the most common Spotify genre label among the artists in each cluster. In few cases, clusters were comprised of artists with no associated genre labels (these appear as ``UNKNOWN" in Figure~\ref{fig:dendrogram}). 

\begin{figure}[!t]
	\centering
	\includegraphics[width=0.825\linewidth]{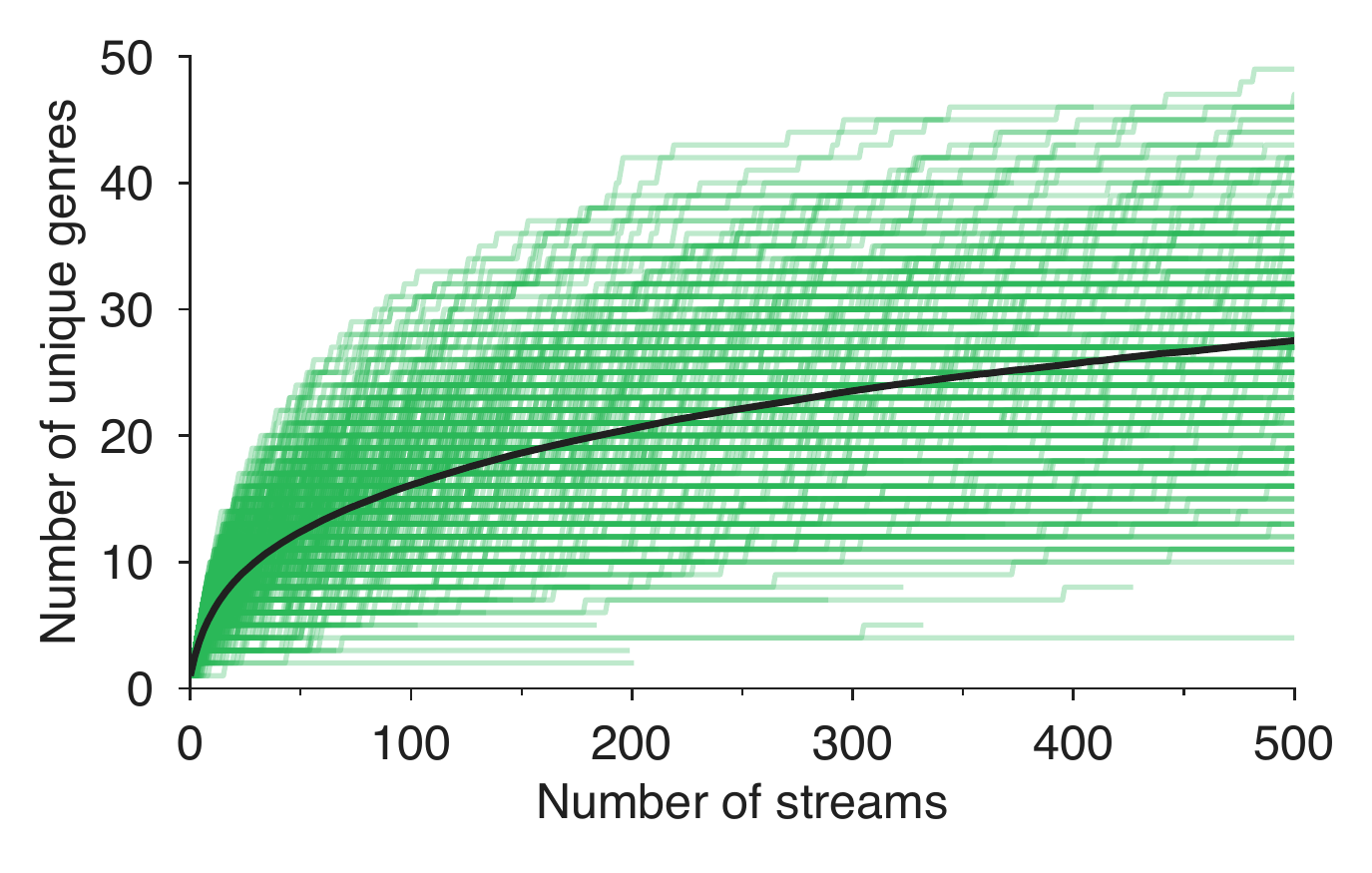}
	\caption{{\bf Rarefaction curves suggest 200 streams capture the range of a listener's musical tastes over our data-derived genres.} {\normalfont Rarefaction curves, shown here for a sample of 500 listeners (average shown in black), indicate that the number of unique genres spanned by each person's streaming history begins to level out after 200 streams. This analysis informed our sampling depth for constructing listeners' taste profiles.}}
	\label{fig:rarefaction}
\end{figure}

In sum, we characterize an individual's musical tastes during each sample period (e.g., $P_1$) by summing together their stream counts for artists in each of the 200 data-derived genres. This process constructs musical taste profiles as 200-dimensional vectors that we analyze using the methods outlined below. To ensure that these vectors are representative of users' tastes, we analyzed rarefaction curves (Figure~\ref{fig:rarefaction}) to determine the minimum number of streams required to construct reliable taste profiles. In the worst case, in which all genres are equally distinct, our analyses suggest a minimum of around 200 streams. This limit informed our sampling depth for each period, ensuring sufficient depth to characterize the tastes of nearly all listeners. Repeating this analysis using the diversity measures introduced below suggests that the range of most users' tastes can typically be inferred from many fewer streams.

\medskip

\textbf{Measuring how tastes change.} Our goal is to quantify the extent to which an individual's musical tastes shift in response to a change in their environment, namely their state of residence. As described above, our construction of musical taste profiles characterizes each person's preferences as a distribution of counts over 200 data-derived genres. Changes in these profiles from one time period to the next can be characterized in two primary directions: (i) changes in \emph{how diverse} a person's tastes are, and (ii) changes in \emph{what genres} a person consumes. Measuring changes of this manner is a central focus in ecological research, which frequently characterizes the diversity \emph{within} (called ``$\alpha$-diversity") or \emph{between} (``$\beta$-diversity") ecosystems~\cite{whittaker1972evolution}, based on counts of species' abundances and measures of their phylogenetic similarity. Here, we draw inspiration from the ecology literature for measuring changes within and between listeners' profiles.

To measure the range or diversity of a person's musical tastes within a given time period, we used Rao-Stirling divergence \cite{rao1982diversity,stirling2007general}. This technique is a popular measure of biodiversity and is closely related to other approaches used throughout ecological research~\cite{martin2002phylogenetic,lozupone2008species}. It has also recently been applied to the study of musical diversity by Park et al.\ ~\cite{park2015understanding}, who highlighted the measure's advantage over existing approaches, namely that some genres can be very similar to others, biasing approaches that count unique genres or otherwise ignore their similarity. Given a taste profile $p$, constructed as a probability distribution over our data-derived genres, Rao-Stirling divergence is calculated as

\begin{equation}
d_{RS}(p) = \sum_{i,j\in K} p_{i} \times p_{j}  \times d(i,j),
\end{equation}

\noindent where $p_{i}$ and $p_{j}$ denote the fraction of streams from genres $i$ and $j$, respectively, and $d(i,j)$ denotes the dissimilarity of the two genres. To quantify the dissimilarity between our data-derived genres, we measured the number of times listeners consumed genres $i$ and $j$ in $P_1$, forming a co-consumption matrix of genres. We then computed $d(i,j)$ as correlation distances, comparing the rows of the resulting matrix (similar to Equation~\ref{eqn:corrdiff}). 

To measure the difference between two taste profiles, we used UniFrac, another approach adopted from the ecology literature. UniFrac is a family of distance metrics used to assess the dissimilarity of two ecosystems that, like Rao-Stirling, takes into account the similarity of the counted elements~\cite{lozupone2005unifrac}. These metrics are constructed using a phylogenetic tree that summarizes the evolutionary distances separating the species or, in our case, the distances between data-derived genres. We used the same genre correlation distances ($d(i,j)$) as in our Rao-Stirling calculations to construct such a tree of genres, using hierarchical clustering (UPGMA algorithm~\cite{sokal1958statistical}; resulting tree shown in Figure~\ref{fig:dendrogram}). Distances between taste profiles were then calculated based on the the amount of distinct versus shared branch length spanned by the two profiles. In our analyses, we used weighted UniFrac ($d_{WU}$), a variant that considers not just which genres are consumed but in what proportions (i.e., the abundance of each lineage).

For completeness, we repeated our UniFrac-based analyses using Jensen-Shannon divergence, a dissimilarity measure that incorporates no information about the relatedness of genres. Jensen-Shannon divergence ($d_{JS}$) is related to the popular Kullback--Leibler ($d_{KL}$) divergence in that $d_{JS}$ is an averaged, symmetrized of measure $d_{KL}$ divergence. Given two probability distributions $a$ and $b$, $d_{KL}$ and $d_{JS}$ are defined as

\begin{align*}
d_{KL}(a||b) & = \sum_{x \in X} a(x) \textrm{log}_2{a(x)\over{b(x)}} \\
d_{JS}(a||b) & = {1\over{2}}d_{KL}(a||c) + {1\over{2}}d_{KL}(b||c),  \\ 
\textrm{where}~c & = {1\over{2}} (a+b). \\
%_{\textrm{this feels unnecessary...}}
\end{align*}

\noindent Qualitatively, our findings were not sensitive to the choice of weighted UniFrac or Jensen-Shannon divergence. As such, we present just the weighted UniFrac-based results below.

\begin{figure*}[!t]
	\centering
	\includegraphics[width=0.91\linewidth]{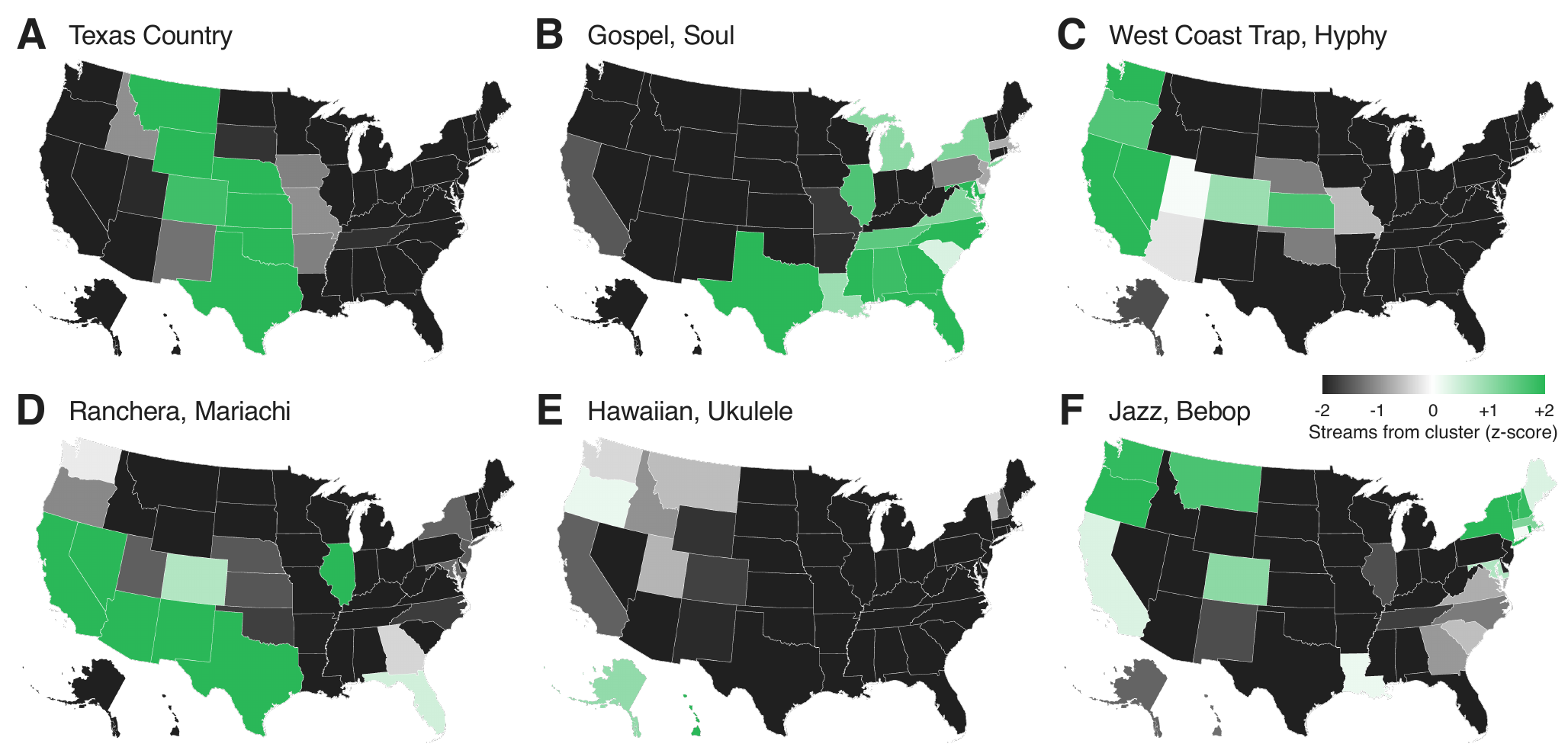}
	\caption{{\bf Data-derived genres encode regional information.} {\normalfont These heatmaps show the fraction of each state's streams coming from six data-derived genres, displayed using $z$-scores. The maps highlight different elements of the states' unique musical identities and provide a useful check to ensure that the genres capture patterns that should be expected historically.}}
	\label{fig:states}
\end{figure*}

\medskip

\textbf{Inferring state-level musical identities.} 
Finally, in several of our analyses, we test whether individuals' musical preferences change in response to relocating from one state to another. To ground these measurements, we constructed state-level musical taste profiles by summing together the streams from all listeners from each state over our 200 data-derived genres. These state-level profiles enable more concrete definitions of change for individuals. For instance, if a person moves from state $a$ to state $b$, do their tastes become more similar to the aggregate profile for state $b$? Or less similar to that of $a$? Additionally, these state-level characterizations provide a valuable sanity check to verify that the states do, in fact, possess distinct musical tastes that might influence the tastes of individuals. Qualitatively, we found that consumption patterns for many data-derived genres matched intuitions based on the history of the states and their ethnic, religious, and cultural compositions (Figure~\ref{fig:states}).

\begin{figure}[!b]
	\centering
	\includegraphics[width=\linewidth]{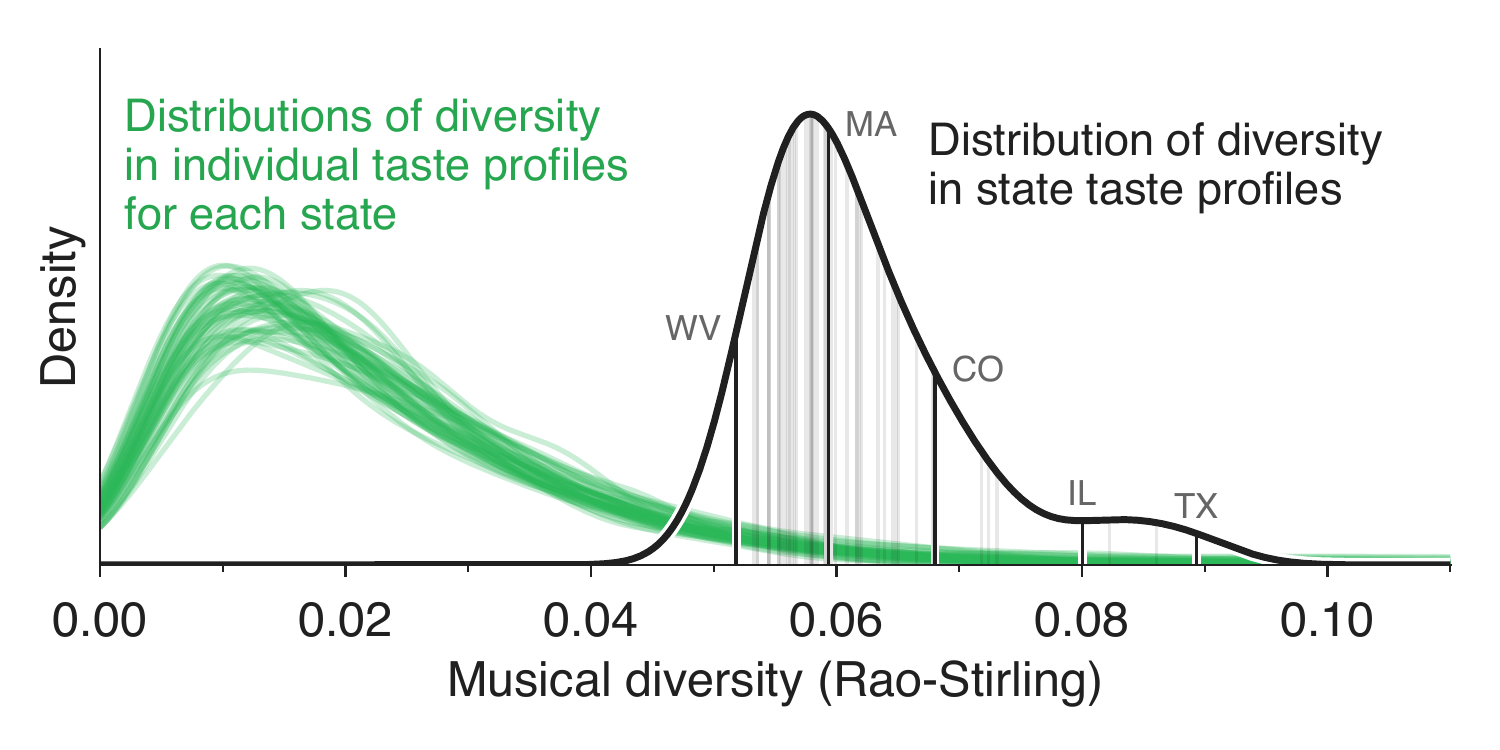}
	\caption{{\bf Musical diversity of individuals follows a similar distribution across the states, even though state-level diversity varies.} {\normalfont Multiple distributions are depicted. In green, we show the distributions of musical taste diversity (Rao-Stirling) for individuals within all fifty states. In black, we show the distributions of musical taste diversity of the aggregate state profiles.}}
	\label{fig:alpha}
\end{figure}

Analyzing the general diversity of these state-level taste profiles using Rao-Stirling divergence, we found that states vary in their aggregate diversity (Figure~\ref{fig:alpha}). This quantity does not correlate with the entropy of U.S.\ Census-reported distributions for race/ethnicity. However, there is a moderate correlation (Pearson's $\rho\!=\!0.49$, $p\!<\!0.001$) between Rao-Stirling diversity and states' Hispanic composition~\cite{kff}. This correlation is likely due in part to large disparities in the co-consumption of genres that differ in the primary languages of their lyrics (see Figure~\ref{fig:dendrogram}), which contributes to higher levels of measured diversity.

While states vary in the diversity of their aggregate compositions, states exhibit similar distributions of diversity calculated at the level of individuals ($p\!>\!0.05$, Conover post-hoc test for multiple comparisons). Together, these two observations suggest that the higher diversity of some state taste profiles is driven by diverse compositions of individuals, not because the individuals in those states have more diverse tastes themselves. 

%*.*.*%.*.*.%
% RESULTS 
%.*.*.%*.*.*%
\section{Results}

We devised two sets of matched pair analyses to study the short- and long-term effects of relocation on individuals' musical tastes. We begin by analyzing short- then long-term effects, followed by an examination of the relationship between the ages of listeners and the music they consume.

\medskip

\textbf{Short-term effects of relocation.} We tested whether moving from one state ($a$) to another ($b$) induces short-term changes in individuals' musical tastes by constructing matched pairs of listeners. Each mover $m$ in our sample frame was paired to a non-mover $n$ by exact matching under the following criteria:

\begin{enumerate}
\item Both $m$ and $n$ lived in state $a$ during $P_1$
\item $m$ moved to state $b$ between $P_1$ and $P_2$
\item $n$ continued living in $a$ during $P_2$
\item $m$ and $n$ share the same reported gender and age group
\end{enumerate}

These criteria formed a total of \mbox{$N\!=\!592,716$} matched pairs of listeners. First, we tested whether movers exhibit a systematic change in the overall diversity of their musical tastes, measured using Rao-Stirling divergence. Specifically, we measured $m$'s change in diversity from before ($P_1$) to after their relocation ($P_2$) as $d_{RS}(m_{P_2}) - d_{RS}(m_{P_1})$. We then compared this change in diversity to the same quantity calculated for $m$'s matched pair individual, $n$. Sampling 1000 matched pairs from each state, we found no significant change in movers' overall taste diversity compared to non-movers, neither between $P_1$ and $P_2$ ($p\!=\!0.72$, matched-pair $t$-test) nor $P_1$ and $P_3$ ($p$=0.24). 

Next, we tested whether movers' taste profiles shift in response to relocation, possibly becoming less similar to their former home and more similar to their new home. First, we calculated the difference in dissimilarities between $m$ and $a$'s taste profiles, during $P_1$ and $P_2$,
\begin{equation}
\label{eqn:diff1}
D(m, a | P_2, P_1) = d_{WU}(m_{P_2}, a_{P_2})\!-\!d_{WU}(m_{P_1}, a_{P_1}).
\end{equation}
This quantity captures whether $m$ is more similar to $a$ during time period $P_1$ or $P_2$. Next, we calculated the same quantity for $n$ to compare, 
\begin{equation}
\label{eqn:diff2}
D(n, a | P_2, P_1) = d_{WU}(n_{P_2}, a_{P_2})\!-\!d_{WU}(n_{P_1}, a_{P_1}).
\end{equation}
The difference between these two quantities, $D(m, a | P_2, P_1) $ and $D(n, a | P_2, P_1)$, captures whether $m$'s dissimilarity to their former home state $a$ increases or decreases after moving to state $b$, compared to their matched pair, who remained in state $a$. Sampling 1000 matched pairs from each state, we found no significant differences in the similarity of states' musical taste profiles to individuals who live in versus moved away from that state ($p\!=\!0.70$, matched-pair $t$-test). That is, moving to a new state does not, in the short term, appear to have a significant effect on individuals' musical tastes, neither making them more nor less like their former home. 

\begin{figure}[!t]
	\centering
	\includegraphics[width=\linewidth]{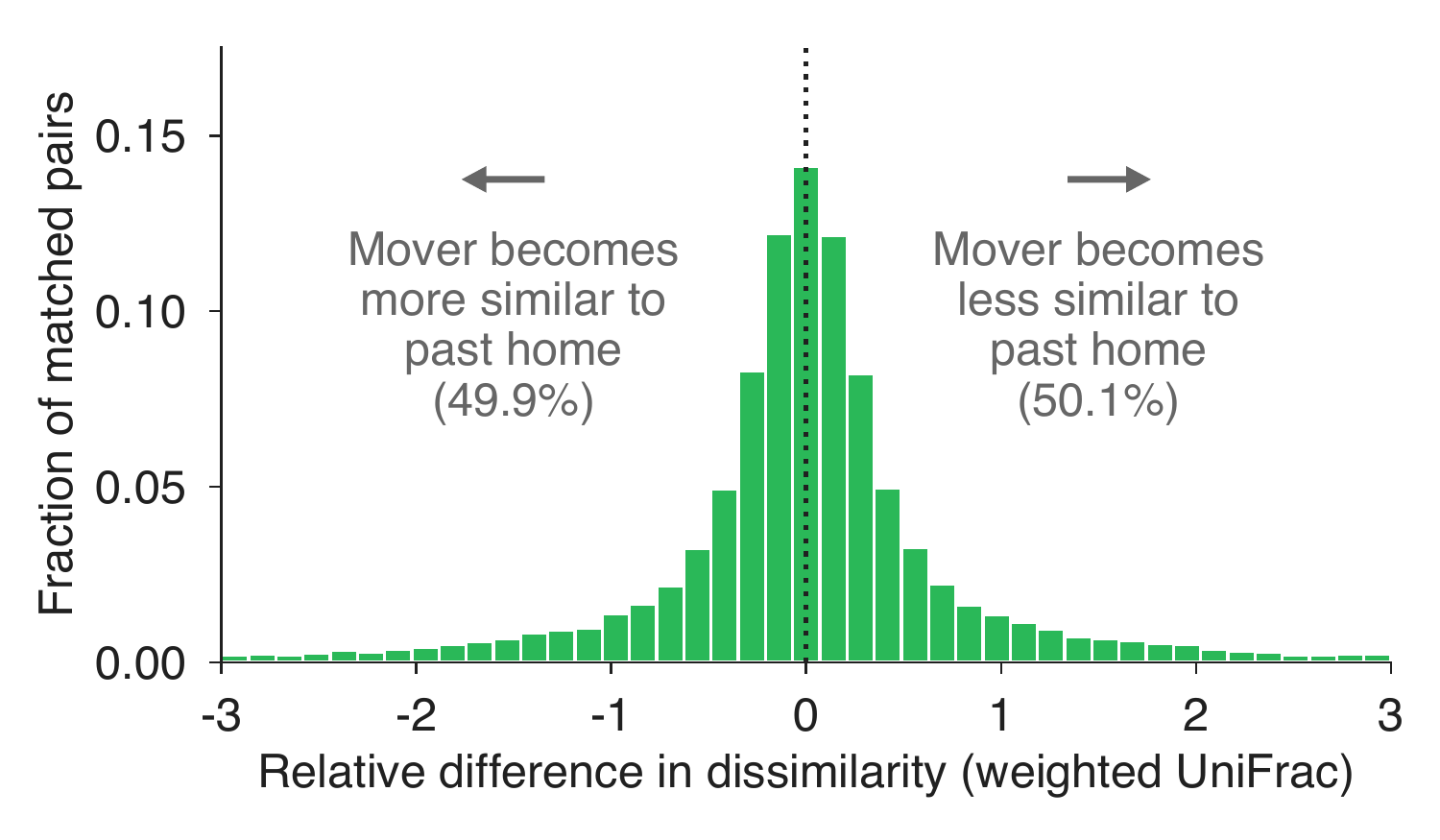}
		\caption{{\bf Short-term effects of relocation on musical tastes are insignificant, within the range of typical month-to-month variability.} {\normalfont The distribution of changes in dissimilarity between movers ($m$) and their previous home state ($a$) compared to non-movers. Changes here are scaled by the amount of expected variability from within-person, month-to-month fluctuations in musical tastes (see main text). As shown, no systematic changes were found across our sample, and individuals' changes were predominantly within the range of typical variability.}}
			\label{fig:mp}
\end{figure}

Dividing these differences by the month-to-month variability of all non-movers (i.e., standard deviation of $D(n, a | P_2, P_1)  - D(n, a | P_3, P_2)$), we note that, despite some amount of heterogeneity in the magnitude and size of individuals' changes, the observed differences generally fall within one standard deviation of typical month-to-month fluctuations in a person's tastes (Figure~\ref{fig:mp}).

Next, replacing $a$ with $b$ in Equations~\ref{eqn:diff1} and \ref{eqn:diff2}, we tested whether individuals' tastes become more or less similar to their new state ($b$) after relocating there. As before, we sampled 1000 matched pairs from each state and found no significant differences between movers and non-movers ($p\!=\!0.77$, matched-pair $t$-test). In the short term, moving to a new state does not seem to move an individual's musical taste profile closer to their new state's. 

The results of these comparisons hold over slightly longer periods of time (i.e.\ comparing taste profiles between $P_1$ and $P_3$, rather than $P_1$ and $P_2$), despite a trend towards significance. They are also robust across different age groups and genders, as well as isolating the effects for particular choices of $a$ or $b$, and using more narrowly-defined state taste profiles, constructed for each age-gender group. In addition, we measured changes relative to another non-mover, rather than aggregate state taste profiles, and found similar outcomes. Lastly, we added another matching criterion to the list above, requiring that matched pairs share the same ``favorite" (i.e., most consumed) genre during $P_1$. Each of these modifications only served to corroborate the observations above.

Our results here indicate that relocation has little effect on individuals' taste profiles in the months following their move. While no detectable changes were found, our measurements cannot rule out the possibility that differences may become significant over longer periods of time. Accordingly, in the next section, we outline a similar matched pair analysis to reason about how relocation affects taste profiles in the long term.

\medskip

\textbf{Long-term effects of relocation.} We tested whether relocation induces long-term changes in individuals' musical tastes by constructing matched pairs of listeners based on their locations during the three home holidays spanned by our sample frame. Listeners who traveled to a different state for at least two of these holidays were assumed to have formerly resided in that state, having since moved to their current state. As in the analysis above, we paired each mover, $m$, with a non-mover $n$---someone who spent the home holidays in their current home state---by applying the following criteria:

\begin{enumerate}
\item $m$ traveled to and is assumed to have lived in state $a$
\item $n$ lives in and spent the holidays in state $a$
\item $m$ now lives in state $b$
\item $m$ and $n$ share the same reported gender and age group
\end{enumerate}

These criteria formed a total of \mbox{$N\!=\!469,935$} matched pairs. First, we tested whether movers differ from non-movers in the general diversity of their musical tastes, measured by Rao-Stirling divergence for aggregate profiles of listeners' streams from periods $P_1$ through $P_3$. Sampling 1000 matched pairs from each state, we found that movers and non-movers had similarly diverse musical taste profiles ($p\!=\!0.41$, Mann-Whitney). This result is unintuitive given that exposure to a wider variety of environments could plausibly instill more varied musical tastes. To ensure that this result was not driven by moves to neighboring states with similar cultures, we repeated this test, omitting moves between states that share a border. Under this restriction, movers exhibit only marginally higher diversity than non-movers (0.1\% higher; $p\!<\!0.001$, matched-pair $t$-test). 

\begin{figure}[!t]
	\centering
	\includegraphics[width=\linewidth]{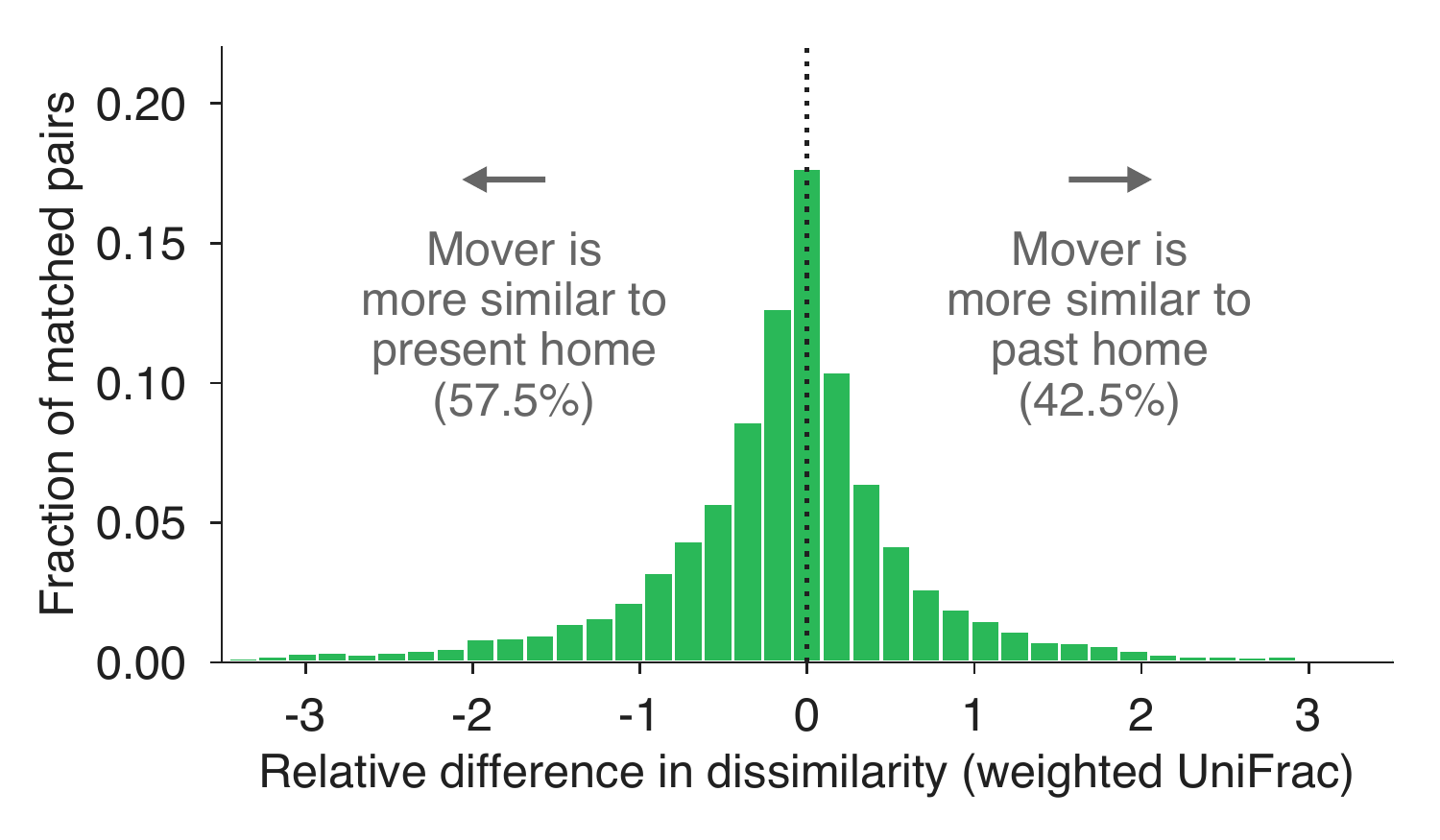}
		\caption{{\bf Long-term effects of relocation on musical tastes indicate that listeners' tastes shift towards their current environments', if only somewhat.} {\normalfont Here, negative values indicate that $m$ is more similar to their present home state ($b$) than their past home state ($a$), relative to $n$. Changes are normalized by the amount of expected variability in non-movers' taste profiles.}}
			\label{fig:mpx}
\end{figure}

Next, using these same aggregate profiles, we determined whether movers' tastes are more similar to their inferred past ($a$) or present ($b$) home states, relative to non-movers. Said differently: do movers' tastes shift to reflect their new environment? To test this possibility, we used weighted UniFrac to compute the difference in dissimilarity between $m$ and the two states as
\begin{equation}
\label{eqn:xmasdiff}
D(m, a, b) = d_{WU}(m, b) - d_{WU}(m, a).
\end{equation}
Again drawing a sample of 1000 matched pairs from each state, we found that the distribution of $D(m, a, b)$ skews positive ($p\!<\!0.001$, $t$-test), with 64\% of movers being more similar to their past home $a$ than their current home $b$. 
Next, we calculated a similar quantity ($D(n, a, b)$) for $m$'s matched pair, $n$, and considered the difference between the resulting quantities. We found that movers are significantly more similar to their present home state than their past home state, compared to their matched pair ($p\!<\!0.001$, matched-pair $t$-test; see Figure~\ref{fig:mpx}). Here, 57.5\% of movers were more similar to their current state.

Together these findings paint a nuanced picture of the long-term effects of relocation: on average, listeners' musical tastes continue to resemble their past home states', and shift only slightly, if at all, towards their present home state. Unlike our first analysis, here we lack information about when listeners may have moved away from their previous home state and therefore how these effects may develop over time. Nevertheless, we find that both past and present environments do appear to shape listener preferences in the long term. To gain a better understanding of when listener environments likely impact musical tastes, we now shift our focus towards the relationship between the ages of listeners and the music they consume.

\medskip

\textbf{Timing of environmental influence.} In the previous section, we found that both past and present environments play a role in shaping listeners' musical tastes. Here, we test when and therefore which environments are likely to affect these tastes by analyzing how listener age predicts the age of the music they consume. Specifically, we look for indications of when musical tastes are formed in order to inform when environment is most likely to have an impact.

\begin{figure}[!t]
	\centering
	\includegraphics[width=0.85\linewidth]{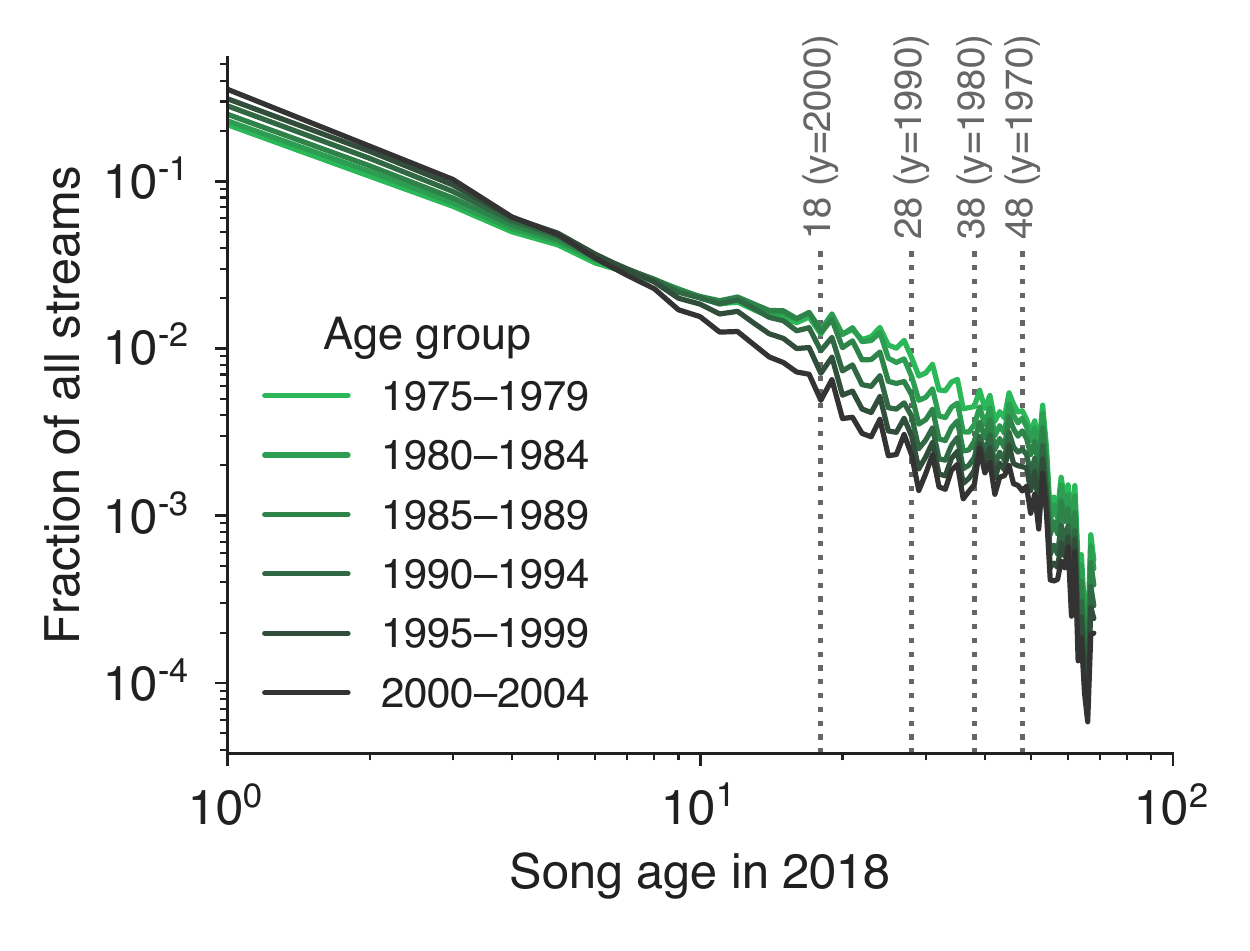}
		\caption{{\bf Listeners of all ages predominantly stream current music.} {\normalfont Distributions of song age (i.e., 2018 minus release year) consumed by our six age groups. Shown on log-log axes, two patterns emerge: (1) regardless of their age, listeners generally consume new or recently-released music, and (2) listener age correlates with song age.}}
		\label{fig:agesraw}
\end{figure}

First, we analyzed the general relationship between the age of listeners and their music (Figure~\ref{fig:agesraw}). We found that listeners of all ages consume predominantly current music, with 28\% of all streams coming from songs that are less than a year old. This observation applies to all age groups, though older listeners are more likely to consume older music.

\begin{figure}[!t]
	\centering
	\includegraphics[width=\linewidth]{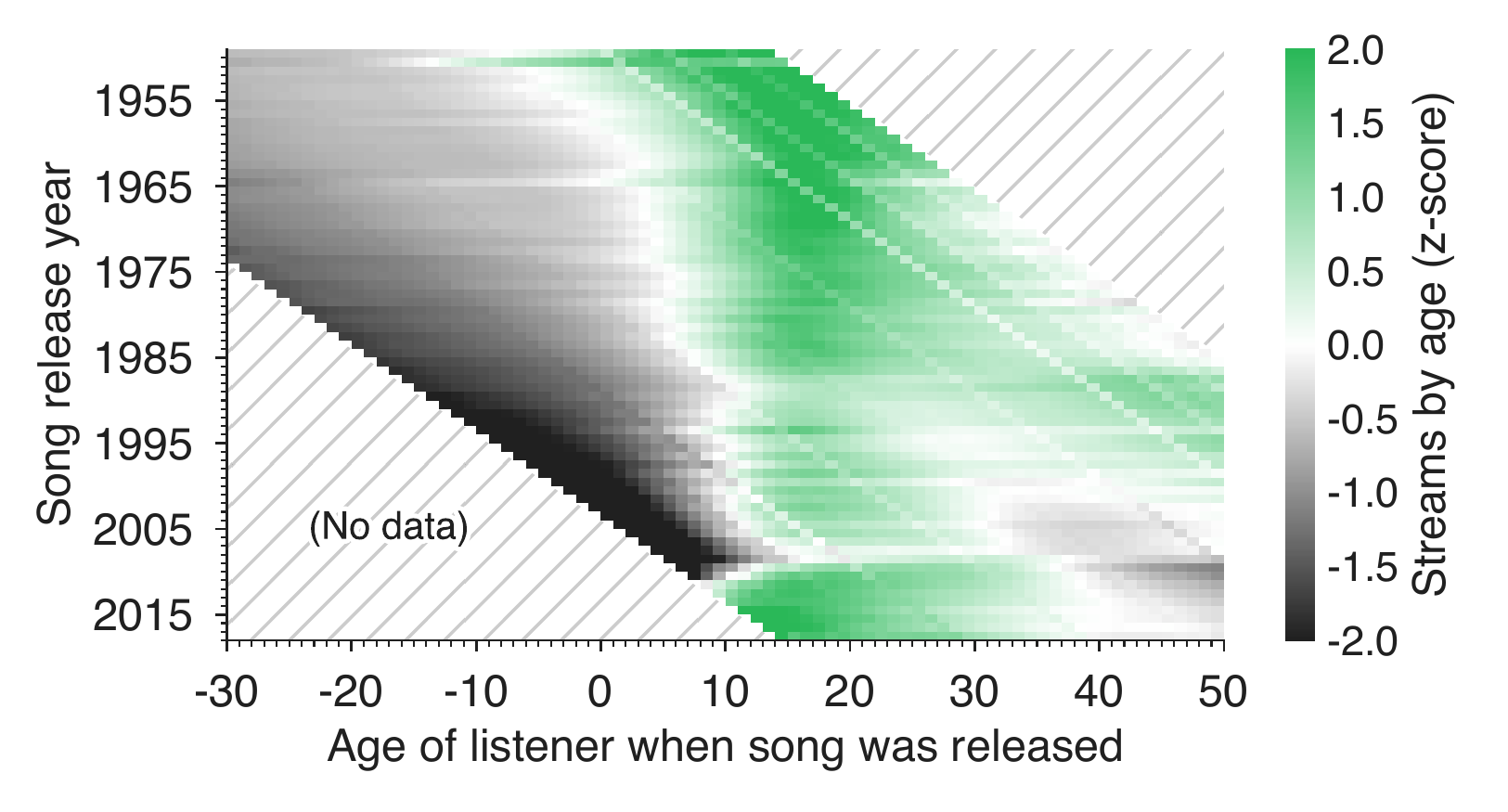}
		\caption{{\bf The distribution of song vs.\ listener age highlights the importance of adolescence in musical identity formation.} {\normalfont For each song release year, the distribution of current listener ages when the song was released. Negative ages indicate listeners streaming music that was released before they were born. Two age regions were excluded due to low representation of older listeners (upper right) and young or unborn listeners (bottom left).}}
	\label{fig:ages}
\end{figure}

While all listeners tend to consume music that has been released recently, music of a given age may be more or less likely to be consumed by listeners of different ages. We analyzed the relationship between a song's release year and the age of its current listeners when the song was released by calculating the distribution of streams by listener age for each song release year. We then calculated z-scores for the fractions of streams from each listener age. Applying this transformation highlights an affinity in listeners for music that was released when they were 10--20 years old (Figure~\ref{fig:ages}).

This pattern corroborates past studies~\cite{ssd2018,Schwartz2003} that similarly found adolescence to be a crucial period in the development of musical taste and identity. In the context of our other results---that musical tastes are generally robust to change but reflect our past locations---this pattern implies that it is both the timing and geographic location of person's adolescence that casts their musical identity.

%*.*.*%.*.*.%*.*.*%
% DISCUSSION 
%.*.*.%*.*.*%.*.*.%
\section{Discussion}
In this study, we used a comprehensive data set on music consumption in the United States to measure the impact of geographic relocation on individuals' musical taste profiles. Analyzing short-term effects, we found that listeners' musical tastes are robust to changes in environment, both in terms of their overall diversity as well as in composition. Over longer periods of time, relocation does appear to shift individuals' tastes marginally towards those of their new environment. The size of this effect, however, is small, and listeners' tend to more strongly resemble their past rather than present environments. Finally, listeners' affinities for music released during their adolescence suggests that a person's musical environment during this period ultimately shapes their musical identity. 

Our results indicate that musical tastes, characterized here as distributions over 200 data-derived genres, are largely robust to relocation from one state to another in the U.S. There are at least three factors that might help explain this observation. First, our analysis studies the changes in listeners' consumption of general styles of music, as captured by the 10,000 most popular artists in the United States. Naturally, the popularity of these artists (and the genres to which they belong) varies tremendously, with the most popular artists receiving orders of magnitude more streams than the least popular artists. We made no effort to re-weight or otherwise adjust for common tastes so as to amplify any differences, yet it may well be the case that artists on the long tail of the popularity distribution are what make listeners' tastes unique~~\cite{anderson2006long,goel2010anatomy}. Our decision to not re-weight taste profiles was made consciously, under the assumption that listeners themselves would define their musical tastes based on how often they listen to each genre, not how unique those genres are. This assumption should itself be explored by future studies in order to better understand how people describe their interests and how descriptions vary depending on social context and the curated version of self the person wishes to portray.

Second, our classification of environment as states of residence masks a large amount of cultural variation within most states, adding noise to our study and potentially hiding subtle patterns. This limitation is particularly true for states with both sizable urban and rural populations. For example, the cultural differences between moving to Manhattan versus a small town in upstate New York can be substantial. We attempted to mitigate this effect in our matched pair analyses by incorporating additional matching criteria (e.g., requiring that matches share favorite genres), but future studies should consider investigating these sub-state cultural variations more directly. For instance, what is the impact of moving from an urban to a rural environment, or vice versa? And, are there personal attributes that predict the malleability of someone's tastes?

Lastly, individuals who use Spotify or similar services may have more robust musical tastes than others. For one, these listeners may have a higher baseline awareness of or interest in music. Perhaps more importantly, however, the success of music streaming platforms stems in part from their ability to give listeners access their music at any time, anywhere. Paired with increasingly prevalent mobile phone technology, these platforms have ushered in a new era of accessibility in music and have accelerated a transition from what has traditionally been a ``push" model of music consumption (i.e., radio stations decide what gets played) to more of a ``pull" model (i.e., individuals decide for themselves what to play), giving listeners more control over what they listen to, when, and how. This increased level of individualized play control may contribute to the portability and thus robustness of tastes observed here. That is, had this analysis been somehow possible 30 years ago, we might expect to see greater malleability of musical tastes because listeners' choice was more directly constrained by what was locally available after a move. Moreover, the effect of this shift in control on the formation of musical tastes represents another interesting direction for future research, to make sense of preferences in light of nearly limitless choice and control~\cite{vanderbilt2016you}. 

Our study focuses on Spotify users in the U.S. in particular, a population that skews towards younger and, by virtue of having access to the Internet and services like Spotify, likely wealthier individuals. This limitation is compounded in our analysis of the long-term effects of relocation, in which we assume past environments are suggested by holiday travel patterns. We acknowledge that these assumptions likely exclude people of lower socioeconomic status, and may introduce some noise, for example, from individuals who regularly travel during the holidays but to a location other than their former home. Other, more precise methods for inferring location histories may be possible and would enable more nuanced analyses in the future. Nevertheless, listeners of higher socioeconomic status are generally regarded as ``cultural omnivores"~\cite{peterson1996changing,park2015understanding}, which may make their tastes more malleable than others and lead us to underestimate the rigidity of musical preferences in the larger population.

Selection biases also complicate the idea of carrying out similar analyses for international migration: relocating between countries is expensive, both financially and socially, and people who do may not be wholly representative of those with more restricted mobility. However, overcoming this limitation would offer valuable insight into the robustness and transmissibility of international culture. For example, what aspects of a culture (e.g., Hofstede's cultural dimensions~\cite{hofstede1984cultural}) make it more or less resistant to change? To what extent do migrants adopt the culture of their new environment, and at what rate? And, could any of these variables be predicted beforehand? As access to services like Spotify increases in the future, answering such questions may become possible.

As people increasingly discover, consume, and share music through online platforms, the field of musicology is uniquely poised to produce new insights on the development of tastes and identity, their determinants, and their interaction with surrounding communities and cultures. In addition to providing insights into these topics, further research in this direction may enhance the way people experience music through these platforms, while also ensuring that services are mindful of their potential impacts on the development of individual identity and on the evolution of culture more broadly.

\section{Acknowledgments}
We thank Manish Nag, Nathan Stein, Scott Wolf, Rozmin Daya, Clay Gibson, Will Shapiro, Glenn McDonald, Ricardo Monti, Laura Norris, Daniel Larremore, Abigail Jacobs, Allison Morgan, and Herrissa Lamothe for helpful conversations.

%\bibliographystyle{apsrev4-1}
%\bibliography{manual_refs}

\begin{thebibliography}{0}%
\makeatletter
\providecommand \@ifxundefined [1]{%
 \@ifx{#1\undefined}
}%
\providecommand \@ifnum [1]{%
 \ifnum #1\expandafter \@firstoftwo
 \else \expandafter \@secondoftwo
 \fi
}%
\providecommand \@ifx [1]{%
 \ifx #1\expandafter \@firstoftwo
 \else \expandafter \@secondoftwo
 \fi
}%
\providecommand \natexlab [1]{#1}%
\providecommand \enquote  [1]{``#1''}%
\providecommand \bibnamefont  [1]{#1}%
\providecommand \bibfnamefont [1]{#1}%
\providecommand \citenamefont [1]{#1}%
\providecommand \href@noop [0]{\@secondoftwo}%
\providecommand \href [0]{\begingroup \@sanitize@url \@href}%
\providecommand \@href[1]{\@@startlink{#1}\@@href}%
\providecommand \@@href[1]{\endgroup#1\@@endlink}%
\providecommand \@sanitize@url [0]{\catcode `\\12\catcode `\$12\catcode
  `\&12\catcode `\#12\catcode `\^12\catcode `\_12\catcode `\%12\relax}%
\providecommand \@@startlink[1]{}%
\providecommand \@@endlink[0]{}%
\providecommand \url  [0]{\begingroup\@sanitize@url \@url }%
\providecommand \@url [1]{\endgroup\@href {#1}{\urlprefix }}%
\providecommand \urlprefix  [0]{URL }%
\providecommand \Eprint [0]{\href }%
\providecommand \doibase [0]{http://dx.doi.org/}%
\providecommand \selectlanguage [0]{\@gobble}%
\providecommand \bibinfo  [0]{\@secondoftwo}%
\providecommand \bibfield  [0]{\@secondoftwo}%
\providecommand \translation [1]{[#1]}%
\providecommand \BibitemOpen [0]{}%
\providecommand \bibitemStop [0]{}%
\providecommand \bibitemNoStop [0]{.\EOS\space}%
\providecommand \EOS [0]{\spacefactor3000\relax}%
\providecommand \BibitemShut  [1]{\csname bibitem#1\endcsname}%
\let\auto@bib@innerbib\@empty
%</preamble>
\end{thebibliography}%


\begin{thebibliography}{47}%
\makeatletter
\providecommand \@ifxundefined [1]{%
 \@ifx{#1\undefined}
}%
\providecommand \@ifnum [1]{%
 \ifnum #1\expandafter \@firstoftwo
 \else \expandafter \@secondoftwo
 \fi
}%
\providecommand \@ifx [1]{%
 \ifx #1\expandafter \@firstoftwo
 \else \expandafter \@secondoftwo
 \fi
}%
\providecommand \natexlab [1]{#1}%
\providecommand \enquote  [1]{``#1''}%
\providecommand \bibnamefont  [1]{#1}%
\providecommand \bibfnamefont [1]{#1}%
\providecommand \citenamefont [1]{#1}%
\providecommand \href@noop [0]{\@secondoftwo}%
\providecommand \href [0]{\begingroup \@sanitize@url \@href}%
\providecommand \@href[1]{\@@startlink{#1}\@@href}%
\providecommand \@@href[1]{\endgroup#1\@@endlink}%
\providecommand \@sanitize@url [0]{\catcode `\\12\catcode `\$12\catcode
  `\&12\catcode `\#12\catcode `\^12\catcode `\_12\catcode `\%12\relax}%
\providecommand \@@startlink[1]{}%
\providecommand \@@endlink[0]{}%
\providecommand \url  [0]{\begingroup\@sanitize@url \@url }%
\providecommand \@url [1]{\endgroup\@href {#1}{\urlprefix }}%
\providecommand \urlprefix  [0]{URL }%
\providecommand \Eprint [0]{\href }%
\providecommand \doibase [0]{http://dx.doi.org/}%
\providecommand \selectlanguage [0]{\@gobble}%
\providecommand \bibinfo  [0]{\@secondoftwo}%
\providecommand \bibfield  [0]{\@secondoftwo}%
\providecommand \translation [1]{[#1]}%
\providecommand \BibitemOpen [0]{}%
\providecommand \bibitemStop [0]{}%
\providecommand \bibitemNoStop [0]{.\EOS\space}%
\providecommand \EOS [0]{\spacefactor3000\relax}%
\providecommand \BibitemShut  [1]{\csname bibitem#1\endcsname}%
\let\auto@bib@innerbib\@empty
%</preamble>
\bibitem [{\citenamefont {DeNora}(2000)}]{denora2000music}%
  \BibitemOpen
  \bibfield  {author} {\bibinfo {author} {\bibfnamefont {Tia}\ \bibnamefont
  {DeNora}},\ }\href@noop {} {\emph {\bibinfo {title} {Music in everyday
  life}}}\ (\bibinfo  {publisher} {Cambridge University Press},\ \bibinfo
  {year} {2000})\BibitemShut {NoStop}%
\bibitem [{\citenamefont {MacDonald}\ \emph {et~al.}(2002)\citenamefont
  {MacDonald}, \citenamefont {Hargreaves},\ and\ \citenamefont
  {Miell}}]{macdonald2002musical}%
  \BibitemOpen
  \bibinfo {editor} {\bibfnamefont {Raymond}\ \bibnamefont {MacDonald}},
  \bibinfo {editor} {\bibfnamefont {David}\ \bibnamefont {Hargreaves}}, \ and\
  \bibinfo {editor} {\bibfnamefont {Dorothy}\ \bibnamefont {Miell}},\ eds.,\
  \href@noop {} {\emph {\bibinfo {title} {Musical identities}}},\ Vol.~\bibinfo
  {volume} {2}\ (\bibinfo  {publisher} {Oxford University Press},\ \bibinfo
  {year} {2002})\BibitemShut {NoStop}%
\bibitem [{\citenamefont {North}\ \emph {et~al.}(2000)\citenamefont {North},
  \citenamefont {Hargreaves},\ and\ \citenamefont
  {O'Neill}}]{north2000importance}%
  \BibitemOpen
  \bibfield  {author} {\bibinfo {author} {\bibfnamefont {Adrian~C}\
  \bibnamefont {North}}, \bibinfo {author} {\bibfnamefont {David~J}\
  \bibnamefont {Hargreaves}}, \ and\ \bibinfo {author} {\bibfnamefont
  {Susan~A}\ \bibnamefont {O'Neill}},\ }\bibfield  {title} {\enquote {\bibinfo
  {title} {The importance of music to adolescents},}\ }\href@noop {} {\bibfield
   {journal} {\bibinfo  {journal} {British Journal of Educational Psychology}\
  }\textbf {\bibinfo {volume} {70}},\ \bibinfo {pages} {255--272} (\bibinfo
  {year} {2000})}\BibitemShut {NoStop}%
\bibitem [{\citenamefont {Schwartz}\ and\ \citenamefont
  {Fouts}(2003)}]{Schwartz2003}%
  \BibitemOpen
  \bibfield  {author} {\bibinfo {author} {\bibfnamefont {Kelly~D.}\
  \bibnamefont {Schwartz}}\ and\ \bibinfo {author} {\bibfnamefont {Gregory~T.}\
  \bibnamefont {Fouts}},\ }\bibfield  {title} {\enquote {\bibinfo {title}
  {{Music preferences, personality style, and developmental issues of
  adolescents}},}\ }\href@noop {} {\bibfield  {journal} {\bibinfo  {journal}
  {Journal of Youth and Adolescence}\ }\textbf {\bibinfo {volume} {32}},\
  \bibinfo {pages} {205--213} (\bibinfo {year} {2003})}\BibitemShut {NoStop}%
\bibitem [{\citenamefont {Sch{\"a}fer}\ and\ \citenamefont
  {Mehlhorn}(2017)}]{SCHAFER2017265}%
  \BibitemOpen
  \bibfield  {author} {\bibinfo {author} {\bibfnamefont {Thomas}\ \bibnamefont
  {Sch{\"a}fer}}\ and\ \bibinfo {author} {\bibfnamefont {Claudia}\ \bibnamefont
  {Mehlhorn}},\ }\bibfield  {title} {\enquote {\bibinfo {title} {Can
  personality traits predict musical style preferences? {A} meta-analysis},}\
  }\href@noop {} {\bibfield  {journal} {\bibinfo  {journal} {Personality and
  Individual Differences}\ }\textbf {\bibinfo {volume} {116}},\ \bibinfo
  {pages} {265 -- 273} (\bibinfo {year} {2017})}\BibitemShut {NoStop}%
\bibitem [{\citenamefont {Sloboda}\ and\ \citenamefont
  {O'neill}(2001)}]{sloboda2001emotions}%
  \BibitemOpen
  \bibfield  {author} {\bibinfo {author} {\bibfnamefont {John~A}\ \bibnamefont
  {Sloboda}}\ and\ \bibinfo {author} {\bibfnamefont {Susan~A}\ \bibnamefont
  {O'neill}},\ }\bibfield  {title} {\enquote {\bibinfo {title} {Emotions in
  everyday listening to music},}\ }\href@noop {} {\bibfield  {journal}
  {\bibinfo  {journal} {Music and emotion: Theory and research}\ ,\ \bibinfo
  {pages} {415--429}} (\bibinfo {year} {2001})}\BibitemShut {NoStop}%
\bibitem [{\citenamefont {Saarikallio}\ and\ \citenamefont
  {Erkkil{\"a}}(2007)}]{saarikallio2007role}%
  \BibitemOpen
  \bibfield  {author} {\bibinfo {author} {\bibfnamefont {Suvi}\ \bibnamefont
  {Saarikallio}}\ and\ \bibinfo {author} {\bibfnamefont {Jaakko}\ \bibnamefont
  {Erkkil{\"a}}},\ }\bibfield  {title} {\enquote {\bibinfo {title} {The role of
  music in adolescents' mood regulation},}\ }\href@noop {} {\bibfield
  {journal} {\bibinfo  {journal} {Psychology of {M}usic}\ }\textbf {\bibinfo
  {volume} {35}},\ \bibinfo {pages} {88--109} (\bibinfo {year}
  {2007})}\BibitemShut {NoStop}%
\bibitem [{\citenamefont {Wells}\ and\ \citenamefont
  {Hakanen}(1991)}]{Wells1991}%
  \BibitemOpen
  \bibfield  {author} {\bibinfo {author} {\bibfnamefont {A}~\bibnamefont
  {Wells}}\ and\ \bibinfo {author} {\bibfnamefont {Ea}~\bibnamefont
  {Hakanen}},\ }\bibfield  {title} {\enquote {\bibinfo {title} {{The emotional
  use of popular-music by adolescents}},}\ }\href@noop {} {\bibfield  {journal}
  {\bibinfo  {journal} {Journalism Quarterly}\ }\textbf {\bibinfo {volume}
  {68}},\ \bibinfo {pages} {445--454} (\bibinfo {year} {1991})}\BibitemShut
  {NoStop}%
\bibitem [{\citenamefont {Erickson}(1996)}]{erickson1996culture}%
  \BibitemOpen
  \bibfield  {author} {\bibinfo {author} {\bibfnamefont {Bonnie~H}\
  \bibnamefont {Erickson}},\ }\bibfield  {title} {\enquote {\bibinfo {title}
  {Culture, class, and connections},}\ }\href@noop {} {\bibfield  {journal}
  {\bibinfo  {journal} {American Journal of Sociology}\ }\textbf {\bibinfo
  {volume} {102}},\ \bibinfo {pages} {217--251} (\bibinfo {year}
  {1996})}\BibitemShut {NoStop}%
\bibitem [{\citenamefont {Bennett}\ and\ \citenamefont
  {Peterson}(2004)}]{bennett2004music}%
  \BibitemOpen
  \bibfield  {author} {\bibinfo {author} {\bibfnamefont {Andy}\ \bibnamefont
  {Bennett}}\ and\ \bibinfo {author} {\bibfnamefont {Richard~A}\ \bibnamefont
  {Peterson}},\ }\href@noop {} {\emph {\bibinfo {title} {Music scenes: local,
  translocal and virtual}}}\ (\bibinfo  {publisher} {Vanderbilt University
  Press},\ \bibinfo {year} {2004})\BibitemShut {NoStop}%
\bibitem [{\citenamefont {Lena}(2012)}]{lena2012banding}%
  \BibitemOpen
  \bibfield  {author} {\bibinfo {author} {\bibfnamefont {Jennifer~C}\
  \bibnamefont {Lena}},\ }\href@noop {} {\emph {\bibinfo {title} {Banding
  together: How communities create genres in popular music}}}\ (\bibinfo
  {publisher} {Princeton University Press},\ \bibinfo {year}
  {2012})\BibitemShut {NoStop}%
\bibitem [{\citenamefont {Cohen}(1991)}]{cohen1991rock}%
  \BibitemOpen
  \bibfield  {author} {\bibinfo {author} {\bibfnamefont {Sara}\ \bibnamefont
  {Cohen}},\ }\href@noop {} {\emph {\bibinfo {title} {Rock culture in
  Liverpool: Popular music in the making}}}\ (\bibinfo  {publisher} {Oxford
  University Press on Demand},\ \bibinfo {year} {1991})\BibitemShut {NoStop}%
\bibitem [{\citenamefont {Hudson}(2006)}]{hudson2006regions}%
  \BibitemOpen
  \bibfield  {author} {\bibinfo {author} {\bibfnamefont {Ray}\ \bibnamefont
  {Hudson}},\ }\bibfield  {title} {\enquote {\bibinfo {title} {Regions and
  place: music, identity and place},}\ }\href@noop {} {\bibfield  {journal}
  {\bibinfo  {journal} {Progress in Human Geography}\ }\textbf {\bibinfo
  {volume} {30}},\ \bibinfo {pages} {626} (\bibinfo {year} {2006})}\BibitemShut
  {NoStop}%
\bibitem [{\citenamefont {Nash}\ and\ \citenamefont
  {Carney}(1996)}]{nash1996seven}%
  \BibitemOpen
  \bibfield  {author} {\bibinfo {author} {\bibfnamefont {Peter~H}\ \bibnamefont
  {Nash}}\ and\ \bibinfo {author} {\bibfnamefont {George~O}\ \bibnamefont
  {Carney}},\ }\bibfield  {title} {\enquote {\bibinfo {title} {The seven themes
  of music geography},}\ }\href@noop {} {\bibfield  {journal} {\bibinfo
  {journal} {Canadian Geographer}\ }\textbf {\bibinfo {volume} {40}},\ \bibinfo
  {pages} {69--74} (\bibinfo {year} {1996})}\BibitemShut {NoStop}%
\bibitem [{\citenamefont {Kong}(1995)}]{kong1995popular}%
  \BibitemOpen
  \bibfield  {author} {\bibinfo {author} {\bibfnamefont {Lily}\ \bibnamefont
  {Kong}},\ }\bibfield  {title} {\enquote {\bibinfo {title} {Popular music in
  geographical analyses},}\ }\href@noop {} {\bibfield  {journal} {\bibinfo
  {journal} {Progress in Human Geography}\ }\textbf {\bibinfo {volume} {19}},\
  \bibinfo {pages} {183--198} (\bibinfo {year} {1995})}\BibitemShut {NoStop}%
\bibitem [{\citenamefont {Carney}(1974)}]{carney1974bluegrass}%
  \BibitemOpen
  \bibfield  {author} {\bibinfo {author} {\bibfnamefont {George~O}\
  \bibnamefont {Carney}},\ }\bibfield  {title} {\enquote {\bibinfo {title}
  {Bluegrass grows all around: The spatial dimensions of a country music
  style},}\ }\href@noop {} {\bibfield  {journal} {\bibinfo  {journal} {Journal
  of Geography}\ }\textbf {\bibinfo {volume} {73}},\ \bibinfo {pages} {34--55}
  (\bibinfo {year} {1974})}\BibitemShut {NoStop}%
\bibitem [{\citenamefont {Hebdige}(2003)}]{hebdige2003cutn}%
  \BibitemOpen
  \bibfield  {author} {\bibinfo {author} {\bibfnamefont {Dick}\ \bibnamefont
  {Hebdige}},\ }\href@noop {} {\emph {\bibinfo {title} {Cutn'Mix: Culture,
  Identity and Caribbean Music}}}\ (\bibinfo  {publisher} {Routledge},\
  \bibinfo {year} {2003})\BibitemShut {NoStop}%
\bibitem [{\citenamefont {Gibson}\ and\ \citenamefont
  {Connell}(2003)}]{gibson2003bongo}%
  \BibitemOpen
  \bibfield  {author} {\bibinfo {author} {\bibfnamefont {Chris}\ \bibnamefont
  {Gibson}}\ and\ \bibinfo {author} {\bibfnamefont {John}\ \bibnamefont
  {Connell}},\ }\bibfield  {title} {\enquote {\bibinfo {title} {{B}ongo {F}ury:
  tourism, music and cultural economy at {B}yron {B}ay, {A}ustralia},}\
  }\href@noop {} {\bibfield  {journal} {\bibinfo  {journal} {Tijdschrift voor
  Economische en Sociale Geografie}\ }\textbf {\bibinfo {volume} {94}},\
  \bibinfo {pages} {164--187} (\bibinfo {year} {2003})}\BibitemShut {NoStop}%
\bibitem [{\citenamefont {Carney}(1998)}]{carney1998music}%
  \BibitemOpen
  \bibfield  {author} {\bibinfo {author} {\bibfnamefont {George}\ \bibnamefont
  {Carney}},\ }\bibfield  {title} {\enquote {\bibinfo {title} {Music
  geography},}\ }\href@noop {} {\bibfield  {journal} {\bibinfo  {journal}
  {Journal of Cultural Geography}\ }\textbf {\bibinfo {volume} {18}},\ \bibinfo
  {pages} {1--10} (\bibinfo {year} {1998})}\BibitemShut {NoStop}%
\bibitem [{\citenamefont {Baily}\ and\ \citenamefont
  {Collyer}(2006)}]{Baily06}%
  \BibitemOpen
  \bibfield  {author} {\bibinfo {author} {\bibfnamefont {John}\ \bibnamefont
  {Baily}}\ and\ \bibinfo {author} {\bibfnamefont {Michael}\ \bibnamefont
  {Collyer}},\ }\bibfield  {title} {\enquote {\bibinfo {title} {Introduction:
  Music and migration},}\ }\href@noop {} {\bibfield  {journal} {\bibinfo
  {journal} {Journal of Ethnic and Migration Studies}\ }\textbf {\bibinfo
  {volume} {32}},\ \bibinfo {pages} {167--182} (\bibinfo {year}
  {2006})}\BibitemShut {NoStop}%
\bibitem [{\citenamefont {{S}potify~{T}echnology {S.A.}}(2018)}]{spotify18}%
  \BibitemOpen
  \bibfield  {author} {\bibinfo {author} {\bibnamefont {{S}potify~{T}echnology
  {S.A.}}},\ }\href@noop {} {\enquote {\bibinfo {title} {Financial results for
  third quarter 2018},}\ } (\bibinfo {year} {2018})\BibitemShut {NoStop}%
\bibitem [{\citenamefont {{IFPI}}(2018)}]{ifpi2018}%
  \BibitemOpen
  \bibfield  {author} {\bibinfo {author} {\bibnamefont {{IFPI}}},\ }\href@noop
  {} {\enquote {\bibinfo {title} {International federation of the phonographic
  industry digital music report 2018},}\ } (\bibinfo {year} {2018})\BibitemShut
  {NoStop}%
\bibitem [{\citenamefont {{US Census Bureau}}(2018)}]{us2018cps}%
  \BibitemOpen
  \bibfield  {author} {\bibinfo {author} {\bibnamefont {{US Census Bureau}}},\
  }\href@noop {} {\enquote {\bibinfo {title} {{CPS} historical
  migration/geographic mobility tables},}\ } (\bibinfo {year}
  {2018})\BibitemShut {NoStop}%
\bibitem [{\citenamefont {Fischer}(1991)}]{fischer1991albion}%
  \BibitemOpen
  \bibfield  {author} {\bibinfo {author} {\bibfnamefont {David~Hackett}\
  \bibnamefont {Fischer}},\ }\href@noop {} {\emph {\bibinfo {title} {Albion's
  seed: Four British folkways in America}}}\ (\bibinfo  {publisher} {Oxford
  University Press},\ \bibinfo {year} {1991})\BibitemShut {NoStop}%
\bibitem [{\citenamefont {Lieske}(1993)}]{lieske1993regional}%
  \BibitemOpen
  \bibfield  {author} {\bibinfo {author} {\bibfnamefont {Joel}\ \bibnamefont
  {Lieske}},\ }\bibfield  {title} {\enquote {\bibinfo {title} {Regional
  subcultures of the united states},}\ }\href@noop {} {\bibfield  {journal}
  {\bibinfo  {journal} {The Journal of Politics}\ }\textbf {\bibinfo {volume}
  {55}},\ \bibinfo {pages} {888--913} (\bibinfo {year} {1993})}\BibitemShut
  {NoStop}%
\bibitem [{\citenamefont {{Allied Van Lines, Inc.}}(2017)}]{Allied2017}%
  \BibitemOpen
  \bibfield  {author} {\bibinfo {author} {\bibnamefont {{Allied Van Lines,
  Inc.}}},\ }\href@noop {} {\enquote {\bibinfo {title} {Why summer is the peak
  moving season},}\ }\bibinfo {howpublished}
  {https://www.allied.com/blog/view/all-blogs/2017/06/20/why-summer-is-the-peak-moving-season}
  (\bibinfo {year} {2017})\BibitemShut {NoStop}%
\bibitem [{\citenamefont {Benney}\ \emph {et~al.}(1959)\citenamefont {Benney},
  \citenamefont {Weiss}, \citenamefont {Meyersohn},\ and\ \citenamefont
  {Riesman}}]{benney1959christmas}%
  \BibitemOpen
  \bibfield  {author} {\bibinfo {author} {\bibfnamefont {Mark}\ \bibnamefont
  {Benney}}, \bibinfo {author} {\bibfnamefont {Robert}\ \bibnamefont {Weiss}},
  \bibinfo {author} {\bibfnamefont {Rolf}\ \bibnamefont {Meyersohn}}, \ and\
  \bibinfo {author} {\bibfnamefont {David}\ \bibnamefont {Riesman}},\
  }\bibfield  {title} {\enquote {\bibinfo {title} {Christmas in an apartment
  hotel},}\ }\href@noop {} {\bibfield  {journal} {\bibinfo  {journal} {American
  Journal of Sociology}\ }\textbf {\bibinfo {volume} {65}},\ \bibinfo {pages}
  {233--240} (\bibinfo {year} {1959})}\BibitemShut {NoStop}%
\bibitem [{\citenamefont {{American Automobile Association}}(2017)}]{aaa2017}%
  \BibitemOpen
  \bibfield  {author} {\bibinfo {author} {\bibnamefont {{American Automobile
  Association}}},\ }\href@noop {} {\enquote {\bibinfo {title} {2017 year-end
  holiday travel forecast review},}\ } (\bibinfo {year} {2017})\BibitemShut
  {NoStop}%
\bibitem [{\citenamefont {Mallett}(2001)}]{mallett2001long}%
  \BibitemOpen
  \bibfield  {author} {\bibinfo {author} {\bibfnamefont {William~J}\
  \bibnamefont {Mallett}},\ }\bibfield  {title} {\enquote {\bibinfo {title}
  {Long-distance travel by low-income households},}\ }\href@noop {} {\bibfield
  {journal} {\bibinfo  {journal} {TRB Transportation Research Circular
  E-C026---Personal Travel: The Long and Short of It}\ ,\ \bibinfo {pages}
  {169--177}} (\bibinfo {year} {2001})}\BibitemShut {NoStop}%
\bibitem [{\citenamefont {Rentfrow}\ and\ \citenamefont
  {Gosling}(2003)}]{rentfrow2003re}%
  \BibitemOpen
  \bibfield  {author} {\bibinfo {author} {\bibfnamefont {Peter~J}\ \bibnamefont
  {Rentfrow}}\ and\ \bibinfo {author} {\bibfnamefont {Samuel~D}\ \bibnamefont
  {Gosling}},\ }\bibfield  {title} {\enquote {\bibinfo {title} {The do re mi's
  of everyday life: the structure and personality correlates of music
  preferences.}}\ }\href@noop {} {\bibfield  {journal} {\bibinfo  {journal}
  {Journal of {P}ersonality and {S}ocial {P}sychology}\ }\textbf {\bibinfo
  {volume} {84}},\ \bibinfo {pages} {1236} (\bibinfo {year}
  {2003})}\BibitemShut {NoStop}%
\bibitem [{\citenamefont {Rentfrow}\ \emph {et~al.}(2011)\citenamefont
  {Rentfrow}, \citenamefont {Goldberg},\ and\ \citenamefont
  {Levitin}}]{Rentfrow:2011aa}%
  \BibitemOpen
  \bibfield  {author} {\bibinfo {author} {\bibfnamefont {Peter~J}\ \bibnamefont
  {Rentfrow}}, \bibinfo {author} {\bibfnamefont {Lewis~R}\ \bibnamefont
  {Goldberg}}, \ and\ \bibinfo {author} {\bibfnamefont {Daniel~J}\ \bibnamefont
  {Levitin}},\ }\bibfield  {title} {\enquote {\bibinfo {title} {The structure
  of musical preferences: a five-factor model},}\ }\href@noop {} {\bibfield
  {journal} {\bibinfo  {journal} {Journal of Personality and Social
  Psychology}\ }\textbf {\bibinfo {volume} {100}},\ \bibinfo {pages}
  {1139--1157} (\bibinfo {year} {2011})}\BibitemShut {NoStop}%
\bibitem [{\citenamefont {Johnston}(2018)}]{spotifygenres18}%
  \BibitemOpen
  \bibfield  {author} {\bibinfo {author} {\bibfnamefont {Maura}\ \bibnamefont
  {Johnston}},\ }\href@noop {} {\enquote {\bibinfo {title} {How {S}potify
  discovers the genres of tomorrow},}\ }\bibinfo {howpublished} {Spotify for
  Artists Blog} (\bibinfo {year} {2018})\BibitemShut {NoStop}%
\bibitem [{\citenamefont {Whittaker}(1972)}]{whittaker1972evolution}%
  \BibitemOpen
  \bibfield  {author} {\bibinfo {author} {\bibfnamefont {Robert~H}\
  \bibnamefont {Whittaker}},\ }\bibfield  {title} {\enquote {\bibinfo {title}
  {Evolution and measurement of species diversity},}\ }\href@noop {} {\bibfield
   {journal} {\bibinfo  {journal} {Taxon}\ ,\ \bibinfo {pages} {213--251}}
  (\bibinfo {year} {1972})}\BibitemShut {NoStop}%
\bibitem [{\citenamefont {Rao}(1982)}]{rao1982diversity}%
  \BibitemOpen
  \bibfield  {author} {\bibinfo {author} {\bibfnamefont {C~Radhakrishna}\
  \bibnamefont {Rao}},\ }\bibfield  {title} {\enquote {\bibinfo {title}
  {Diversity and dissimilarity coefficients: a unified approach},}\ }\href@noop
  {} {\bibfield  {journal} {\bibinfo  {journal} {Theoretical Population
  Biology}\ }\textbf {\bibinfo {volume} {21}},\ \bibinfo {pages} {24--43}
  (\bibinfo {year} {1982})}\BibitemShut {NoStop}%
\bibitem [{\citenamefont {Stirling}(2007)}]{stirling2007general}%
  \BibitemOpen
  \bibfield  {author} {\bibinfo {author} {\bibfnamefont {Andy}\ \bibnamefont
  {Stirling}},\ }\bibfield  {title} {\enquote {\bibinfo {title} {A general
  framework for analysing diversity in science, technology and society},}\
  }\href@noop {} {\bibfield  {journal} {\bibinfo  {journal} {Journal of the
  Royal Society Interface}\ }\textbf {\bibinfo {volume} {4}},\ \bibinfo {pages}
  {707--719} (\bibinfo {year} {2007})}\BibitemShut {NoStop}%
\bibitem [{\citenamefont {Martin}(2002)}]{martin2002phylogenetic}%
  \BibitemOpen
  \bibfield  {author} {\bibinfo {author} {\bibfnamefont {Andrew~P}\
  \bibnamefont {Martin}},\ }\bibfield  {title} {\enquote {\bibinfo {title}
  {Phylogenetic approaches for describing and comparing the diversity of
  microbial communities},}\ }\href@noop {} {\bibfield  {journal} {\bibinfo
  {journal} {Applied and Environmental Microbiology}\ }\textbf {\bibinfo
  {volume} {68}},\ \bibinfo {pages} {3673--3682} (\bibinfo {year}
  {2002})}\BibitemShut {NoStop}%
\bibitem [{\citenamefont {Lozupone}\ and\ \citenamefont
  {Knight}(2008)}]{lozupone2008species}%
  \BibitemOpen
  \bibfield  {author} {\bibinfo {author} {\bibfnamefont {Catherine~A}\
  \bibnamefont {Lozupone}}\ and\ \bibinfo {author} {\bibfnamefont {Rob}\
  \bibnamefont {Knight}},\ }\bibfield  {title} {\enquote {\bibinfo {title}
  {Species divergence and the measurement of microbial diversity},}\
  }\href@noop {} {\bibfield  {journal} {\bibinfo  {journal} {FEMS Microbiology
  Reviews}\ }\textbf {\bibinfo {volume} {32}},\ \bibinfo {pages} {557--578}
  (\bibinfo {year} {2008})}\BibitemShut {NoStop}%
\bibitem [{\citenamefont {Park}\ \emph {et~al.}(2015)\citenamefont {Park},
  \citenamefont {Weber}, \citenamefont {Naaman},\ and\ \citenamefont
  {Vieweg}}]{park2015understanding}%
  \BibitemOpen
  \bibfield  {author} {\bibinfo {author} {\bibfnamefont {Minsu}\ \bibnamefont
  {Park}}, \bibinfo {author} {\bibfnamefont {Ingmar}\ \bibnamefont {Weber}},
  \bibinfo {author} {\bibfnamefont {Mor}\ \bibnamefont {Naaman}}, \ and\
  \bibinfo {author} {\bibfnamefont {Sarah}\ \bibnamefont {Vieweg}},\ }\bibfield
   {title} {\enquote {\bibinfo {title} {Understanding musical diversity via
  online social media},}\ }in\ \href@noop {} {\emph {\bibinfo {booktitle}
  {Ninth International Conference on Web and Social Media}}}\ (\bibinfo {year}
  {2015})\BibitemShut {NoStop}%
\bibitem [{\citenamefont {Lozupone}\ and\ \citenamefont
  {Knight}(2005)}]{lozupone2005unifrac}%
  \BibitemOpen
  \bibfield  {author} {\bibinfo {author} {\bibfnamefont {Catherine}\
  \bibnamefont {Lozupone}}\ and\ \bibinfo {author} {\bibfnamefont {Rob}\
  \bibnamefont {Knight}},\ }\bibfield  {title} {\enquote {\bibinfo {title}
  {Uni{F}rac: a new phylogenetic method for comparing microbial communities},}\
  }\href@noop {} {\bibfield  {journal} {\bibinfo  {journal} {{A}pplied and
  {E}nvironmental {M}icrobiology}\ }\textbf {\bibinfo {volume} {71}},\ \bibinfo
  {pages} {8228--8235} (\bibinfo {year} {2005})}\BibitemShut {NoStop}%
\bibitem [{\citenamefont {Sokal}(1958)}]{sokal1958statistical}%
  \BibitemOpen
  \bibfield  {author} {\bibinfo {author} {\bibfnamefont {Robert~R}\
  \bibnamefont {Sokal}},\ }\bibfield  {title} {\enquote {\bibinfo {title} {A
  statistical method for evaluating systematic relationship},}\ }\href@noop {}
  {\bibfield  {journal} {\bibinfo  {journal} {University of Kansas Science
  Bulletin}\ }\textbf {\bibinfo {volume} {28}},\ \bibinfo {pages} {1409--1438}
  (\bibinfo {year} {1958})}\BibitemShut {NoStop}%
\bibitem [{\citenamefont {{Kaiser Family Foundation}}(2018)}]{kff}%
  \BibitemOpen
  \bibfield  {author} {\bibinfo {author} {\bibnamefont {{Kaiser Family
  Foundation}}},\ }\href@noop {} {\enquote {\bibinfo {title} {Population
  distribution by race/ethnicity},}\ }\bibinfo {howpublished}
  {https://www.kff.org/other/state-indicator/distribution-by-raceethnicity/}
  (\bibinfo {year} {2018})\BibitemShut {NoStop}%
\bibitem [{\citenamefont {Stephens-Davidowitz}(2018)}]{ssd2018}%
  \BibitemOpen
  \bibfield  {author} {\bibinfo {author} {\bibfnamefont {Seth}\ \bibnamefont
  {Stephens-Davidowitz}},\ }\bibfield  {title} {\enquote {\bibinfo {title} {The
  songs that bind},}\ }\href@noop {} {\bibfield  {journal} {\bibinfo  {journal}
  {The New York Times}\ } (\bibinfo {year} {2018})}\BibitemShut {NoStop}%
\bibitem [{\citenamefont {Anderson}(2006)}]{anderson2006long}%
  \BibitemOpen
  \bibfield  {author} {\bibinfo {author} {\bibfnamefont {Chris}\ \bibnamefont
  {Anderson}},\ }\href@noop {} {\emph {\bibinfo {title} {The long tail: Why the
  future of business is selling more for less}}}\ (\bibinfo  {publisher}
  {Hyperion},\ \bibinfo {year} {2006})\BibitemShut {NoStop}%
\bibitem [{\citenamefont {Goel}\ \emph {et~al.}(2010)\citenamefont {Goel},
  \citenamefont {Broder}, \citenamefont {Gabrilovich},\ and\ \citenamefont
  {Pang}}]{goel2010anatomy}%
  \BibitemOpen
  \bibfield  {author} {\bibinfo {author} {\bibfnamefont {Sharad}\ \bibnamefont
  {Goel}}, \bibinfo {author} {\bibfnamefont {Andrei}\ \bibnamefont {Broder}},
  \bibinfo {author} {\bibfnamefont {Evgeniy}\ \bibnamefont {Gabrilovich}}, \
  and\ \bibinfo {author} {\bibfnamefont {Bo}~\bibnamefont {Pang}},\ }\bibfield
  {title} {\enquote {\bibinfo {title} {Anatomy of the long tail: ordinary
  people with extraordinary tastes},}\ }in\ \href@noop {} {\emph {\bibinfo
  {booktitle} {Proceedings of the third ACM International Conference on Web
  Search and Data Mining}}}\ (\bibinfo {organization} {ACM},\ \bibinfo {year}
  {2010})\ pp.\ \bibinfo {pages} {201--210}\BibitemShut {NoStop}%
\bibitem [{\citenamefont {Vanderbilt}(2016)}]{vanderbilt2016you}%
  \BibitemOpen
  \bibfield  {author} {\bibinfo {author} {\bibfnamefont {Tom}\ \bibnamefont
  {Vanderbilt}},\ }\href@noop {} {\emph {\bibinfo {title} {You May Also Like:
  Taste in an Age of Endless Choice}}}\ (\bibinfo  {publisher} {Simon and
  Schuster},\ \bibinfo {year} {2016})\BibitemShut {NoStop}%
\bibitem [{\citenamefont {Peterson}\ and\ \citenamefont
  {Kern}(1996)}]{peterson1996changing}%
  \BibitemOpen
  \bibfield  {author} {\bibinfo {author} {\bibfnamefont {Richard~A}\
  \bibnamefont {Peterson}}\ and\ \bibinfo {author} {\bibfnamefont {Roger~M}\
  \bibnamefont {Kern}},\ }\bibfield  {title} {\enquote {\bibinfo {title}
  {Changing highbrow taste: From snob to omnivore},}\ }\href@noop {} {\bibfield
   {journal} {\bibinfo  {journal} {American Sociological Review}\ ,\ \bibinfo
  {pages} {900--907}} (\bibinfo {year} {1996})}\BibitemShut {NoStop}%
\bibitem [{\citenamefont {Hofstede}(1984)}]{hofstede1984cultural}%
  \BibitemOpen
  \bibfield  {author} {\bibinfo {author} {\bibfnamefont {Geert}\ \bibnamefont
  {Hofstede}},\ }\bibfield  {title} {\enquote {\bibinfo {title} {Cultural
  dimensions in management and planning},}\ }\href@noop {} {\bibfield
  {journal} {\bibinfo  {journal} {Asia Pacific Journal of Management}\ }\textbf
  {\bibinfo {volume} {1}},\ \bibinfo {pages} {81--99} (\bibinfo {year}
  {1984})}\BibitemShut {NoStop}%
\end{thebibliography}


\end{document}